\pdfoutput=1
\documentclass[reprint,aps,fleqn,usenatbib,nofootinbib]{revtex4-1}
\usepackage{graphicx}
\usepackage{amsmath,amssymb}
\usepackage{lastpage}

\usepackage{hyperref}
\usepackage[all]{hypcap}
\usepackage{float}
\usepackage{subfigure}
\usepackage{afterpage}
\usepackage[usenames]{color}
\usepackage{yfonts}
\usepackage{mathrsfs}
\usepackage{upgreek}
\usepackage{gensymb}

\begin{document}

\preprint{APS/123-QED}

\title{Fifth force constraints from galaxy warps}

\author{Harry Desmond}%
 \affiliation{Astrophysics, University of Oxford, Denys Wilkinson Building, Keble Road, Oxford OX1 3RH, UK}
 \email{harry.desmond@physics.ox.ac.uk}
\author{Pedro~G.~Ferreira}%
 \affiliation{Astrophysics, University of Oxford, Denys Wilkinson Building, Keble Road, Oxford OX1 3RH, UK}
\author{Guilhem Lavaux}%
 \affiliation{Sorbonne Universit{\'e}, CNRS, UMR 7095, Institut d'Astrophysique de Paris, 98 bis bd Arago, 75014 Paris, France}
 \affiliation{Sorbonne Universit{\'e}s, Institut Lagrange de Paris (ILP), 98 bis bd Arago, 75014 Paris, France}
\author{Jens Jasche}%
 \affiliation{The Oskar Klein Centre, Department of Physics, Stockholm University, Albanova University Center, SE 106 91 Stockholm, Sweden}
 \affiliation{Excellence Cluster Universe, Technische Universit{\"a}t M{\"u}nchen, Boltzmannstrasse 2, D-85748 Garching, Germany}

\date{\today}

\begin{abstract}
Intra-galaxy signals contain a wealth of information on fundamental physics, both the dark sector and the nature of gravity. While so far largely unexplored, such probes are set to rise dramatically in importance as upcoming surveys provide data of unprecedented quantity and quality on galaxy structure and dynamics. In this paper, we use warping of stellar disks to test the chameleon- or symmetron-screened fifth forces which generically arise when new fields couple to matter. We take $r$-band images of mostly late-type galaxies from the \textit{Nasa Sloan Atlas} and develop an automated algorithm to quantify the degree of U-shaped warping they exhibit. We then forward-model the warp signal as a function of fifth-force strength, $\Delta G/G_N$, and range, $\lambda_C$, and the gravitational environments and internal properties of the galaxies, including full propagation of the non-Gaussian uncertainties. Convolving this fifth-force likelihood function with a Gaussian describing astrophysical and observational noise and then constraining $\Delta G/G_N$ and $\lambda_C$ by Markov Chain Monte Carlo, we find the overall likelihood to be significantly increased ($\Delta\log(\mathcal{L}) \simeq 20$) by adding a screened fifth force with $\lambda_C \simeq 2$ Mpc and $\Delta G/G_N \simeq 0.01$. The variation of $\Delta\log(\mathcal{L})$ with $\lambda_C$ is quantitatively as expected from the correlation of the magnitude of the fifth-force field with the force's range, and a similar model without screening achieves no increase in likelihood over the General Relativistic case $\Delta G=0$. Although these results are in good agreement with a previous analysis of the same model using offsets between galaxies' stellar and gas mass centroids \cite{Desmond_PRD}, we caution that the effects of confounding baryonic and dark matter physics must be thoroughly investigated for the results of the inference to be unambiguous.
\end{abstract}

\maketitle

\section{Introduction}
\label{sec:intro}

Our current theory of fundamental physics is both  successful and puzzling. We are able to explain almost all phenomena at high energies and on cosmological scales in terms of a subset of gauge theories of particles and fields combined with the theory of General Relativity. Yet, while the agreement with data is remarkable, a number of conceptual and practical problems have emerged: why is there a hierarchy of scales, for example, and what is the nature of the Universe's dark sector?

At the heart of almost all attempts to solve these problems are extra degrees of freedom, or new fundamental fields. These emerge when effective field theories break down and when symmetries or conservation laws are broken. The Higgs field is a notable example within the standard model, as is the attempt outside of it to understand inertia by endowing the gravitational constant with dynamics \cite{BD}. They also arise naturally in effective descriptions of cosmology from UV-complete theories. The ubiquity of extra fields in attempts to extend the standard model makes them the smoking gun for new physics. 

New fields generically result in new (``\emph{fifth}'') forces. Furthermore, through quantum consistency at low energy, these fields will couple non-minimally to the gravitational sector, making the forces {\it gravitational} \cite{clifton}. A notable example is a scalar field $\varphi$ with a potential $V(\varphi)\propto m^2\varphi^2+\lambda\varphi^4$; quantum corrections will inevitably lead to a non-minimal coupling of the form $\xi\varphi^2R$ (where $R$ is the Ricci scalar) with $\xi\propto \lambda$ \cite{markkanen}. 

Gravitational fifth forces can be interpreted as arising from corrections to the Newtonian potential $\Phi$. Near a localised body of mass $M$,
\begin{eqnarray}
\Phi_\text{tot}=-\frac{G_N M}{r}-\frac{\Delta G M}{r}e^{-mr}
\end{eqnarray}
where $G_N$ is Newton's constant and $\Delta G$ and $m$ parametrise the relative strength and inverse range of the fifth force. If $m$ is large the force is short-range (i.e. small Compton wavelength $\lambda_C$), but if $m\simeq 0$ the fifth force may compete with gravity on all scales. Stringent constraints have been placed on $\Delta G/G_N$ and $m$ in a variety of regimes, to the point that universally-coupled fifth forces are effectively irrelevant on both laboratory and astrophysical scales \cite{adelberger}.

Over the last decade, however, a better understanding of fifth forces in a cosmological setting has emerged. It has been shown that many theories are equipped with a gravitational {\it screening mechanism}, such that either the range of the fifth force is vastly reduced or its strength becomes negligible in particular environments. For example, in the case of {\it chameleon} \cite{chameleon} or {\it symmetron} \cite{Hinterbichler} screening the fifth forces switches off in regions of high density, as quantified by a deep Newtonian potential; in the case of {\it Vainshtein} screening \cite{vainshtein}, fifth forces are suppressed close to sufficiently massive bodies. Fifth forces may therefore be undetectable in the Solar System and laboratory while playing an important role cosmologically or in regions of space with weak gravitational field.

Screening necessitates that one searches for astrophysical fifth forces in special places, i.e. in particular cosmic environments. The first step, then, is to construct a reliable {\it gravitational map} of a volume of interest. By taking into account the various contributions sourcing the gravitational field, it is possible to demarcate regions of the local Universe by Newtonian potential, acceleration or curvature. In \cite{Desmond} we presented state-of-the-art maps of this type out to $z=0.05$, built from the 2M++ all-sky redshift survey \cite{2M++} combined with the halo population of an N-body simulation \cite{DarkSky} and constrained realisations of the long wavelength modes of the density field \cite{Lavaux}. 

With the ability to identify ``interesting" regions, one can now look for specific signatures of fifth forces. A particularly fertile, and mostly unexplored, regime is within galaxies, where large and growing datasets exist unhampered by cosmic variance. The trade-off is that potential degeneracies arise from non-gravitational (dissipative) galactic physics, as well as uncertainties in the location of the dark matter that governs galaxies' dynamics and drives their evolution. Nevertheless, conservative models of all possible contaminants combined with statistical samples of galaxies in a range of gravitational environments should allow tests of screened fifth forces with existing data more powerful than are possible either within the Solar System or in cosmology.

One such attempt looked for offsets between the centre of emission of the stars (which typically screen themselves) and neutral atomic hydrogen (H\textsc{i}) gas of individual galaxies (\cite{Desmond_PRL} and \cite{Desmond_PRD}, hereafter D18). We demonstrated great sensitivity to $\Delta G /G_N$ and $m$: with strongly conservative measurement errors we set $\Delta G /G_N < {\rm few} \times 10^{-4}$ for fifth-force ranges $\sim 50$ Mpc, while a more realistic noise model led to a 6.6$\sigma$ detection of $\Delta G /G_N \simeq 0.025$ at a range $\lambda_C\simeq1.8$ Mpc. While this result is striking and somewhat intriguing, there may be unaccounted-for systematics that need to be better understood: in particular the differential impact of baryonic physics on stars and gas may be significant, and a proper assessment of this effect awaits study in high-resolution cosmological hydrodynamical simulations of galaxy formation.

We focus here on another signal of fifth forces, warping of stellar galactic disks. Most late-type galaxies appear to be warped to some degree, with the most prominent examples observed in H\textsc{i}. There is weaker evidence for warping in stellar disks, although statistically significant samples have recently begun to be compiled. The origin of warps is not fully understood: they are likely due largely to environmental factors such as tidal interactions or intergalactic magnetic fields, although the observation of warps in isolated galaxies requires that internal processes play some role. For a review of the observational and theoretical status of warps see \cite{Binney}.

Another source of warping is a screened fifth force~\cite{Jain_Vanderplas}. If the (screened) stellar disk lags behind the halo centre in an overall unscreened galaxy, the differential force due to the dark matter across the disk establishes a bowl-shaped equilibrium profile for the stellar mass. This effect is more pronounced the stronger the fifth force and the better its alignment with the disk normal, and is more visible on the plane of the sky the nearer the galaxy is to being viewed edge-on.

In this paper we develop in detail the formalism for warping under a fifth force, and create a likelihood framework with which the properties of the force may thereby be inferred. We then compile and analyse a catalogue of warps in mostly late-type galaxies to constrain $\Delta G/G_N$ and $\lambda_C$. Our analysis parallels in many respects that of D18, although -- and crucially -- its different observational and theoretical inputs give it a largely disjoint set of systematics. Our inference is therefore complementary to D18, and, as we shall see, reinforces its conclusions.

The structure of the paper is as follows. In Section \ref{sec:signal} we model galaxy structure under a fifth force and develop a summary statistic to quantify the overall strength of a U-shaped warp. In Section \ref{sec:data} we describe our observational sample. In Section \ref{sec:method} we detail our method, including measurement of warp strength in the observations, calculation of the predicted fifth-force signal and comparison of the two with a Bayesian likelihood model. In Section \ref{sec:results} we present constraints on $\Delta G/G_N$ and $\lambda_C$ and document a range of checks aimed at validation. Section \ref{sec:discussion} discusses our results, with particular emphasis on potentially confounding systematics, complementarity to D18 and necessary further work. Section \ref{sec:conc} concludes.

\section{Galaxy warps under a fifth force}
\label{sec:signal}

\subsection{Heuristic overview}
\label{sec:heuristic}

The basic requirement for the type of warping we investigate is that a galaxy's stellar disk and dark matter halo experience different accelerations due to their common environment. This may be caused by an additional force besides standard gravity acting on one of the two components, either a long-range interaction with surrounding mass or an additional interaction between the particles comprising one of the two subsystems themselves. Effects of the latter type arise from dynamical friction and dark matter self-interactions; in this paper we focus on the former.

Consider a thin stellar disk at the centre of a dark matter halo, with both mass components accelerating together in an external field. Now suppose that the halo is subject to an additional force which the stars do not feel, so that it experiences an extra acceleration $\vec{a}_5$. The halo centre will move ahead of the stellar centroid until the acceleration of the latter due to its offset $r_*$ from the former exactly compensates for its not feeling the extra force. This sets up a potential gradient across the disk that causes it to warp into a U-shape: heuristically, as the total halo acceleration is lower at larger galactocentric radius it must point at a smaller angle to $\vec{a}_5$ to equal the fixed acceleration difference between the two components (Fig.~\ref{fig:cartoon}). We calculate the expected warp curve precisely in Sec.~\ref{sec:derivation}, and in Sec.~\ref{sec:statistic} we integrate over it with a suitable kernel to create a single measure of overall U-shaped warp strength. We then calculate this quantity analytically for power-law halo density profiles.

The case in which we are particularly interested is a chameleon- or symmetron-screened fifth force \cite{chameleon,Hinterbichler}, characterised by a strength relative to gravity, $\Delta G/G_N$, and a range $\lambda_C$ inversely related to the scalar field's mass. The self-screening parameter $\chi=\phi_0/(2\beta M_\text{pl})$ -- where $\phi_0$ is the background value of the scalar field, $\beta$ is its coupling coefficient to normal matter and $M_\text{pl}$ is the Planck mass -- sets the threshold $|\Phi_c|$ in Newtonian potential above which an object is screened \cite{Zhao_1,Zhao_2,Cabre}. For values of $|\Phi_c|$ below $\sim 10^{-6}$ ($c \equiv 1$), main-sequence stars screen themselves even if the galaxy as a whole is unscreened~\cite{Hui}. Thus the halo feels a fifth force due to surrounding unscreened mass while the stellar disk does not, causing an effective equivalence principle violation of precisely the type necessary to generate the warping described above.

\subsection{Warp curve expected from a screened fifth force}
\label{sec:derivation}

\begin{figure}
  \centering
  \includegraphics[width=0.5\textwidth]{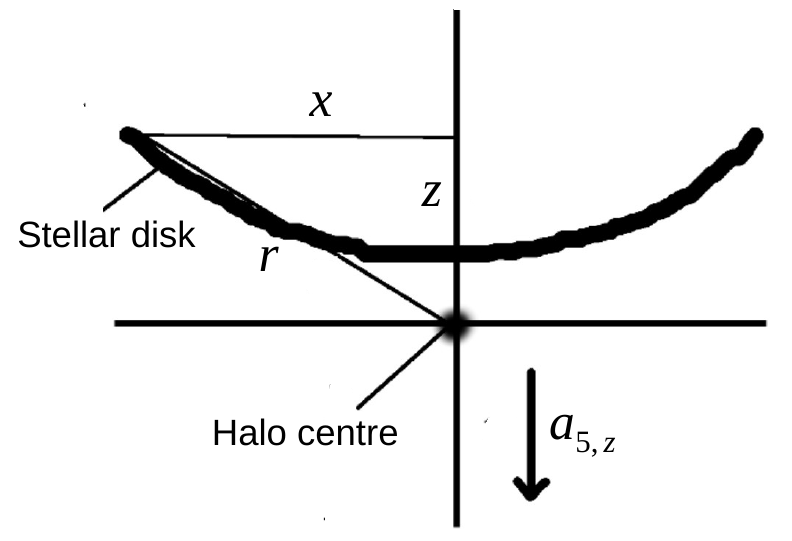}
  \caption{Schematic illustration of the formation of a U-shaped warp from a separation of stellar disk and halo centres. $\vec{a}_5$ is the acceleration resulting from the fifth force, which in the case of thin-shell screening (e.g. chameleon and symmetron) is typically felt by the dark matter but not the stars.}
  \label{fig:cartoon}
\end{figure}

In this section we calculate the shape of the stellar disk resulting from a screened fifth force and investigate its dependence on both fifth-force and galaxy structural parameters. We begin by defining a cylindrical coordinate system with $z$-axis along the disk normal and origin coincident with the halo centre, and consider a star moving in a circular orbit around the $z$-axis (Fig.~\ref{fig:cartoon}). D18 (and previously~\cite{Jain_Vanderplas}) considered the equilibrium separation $r_* \equiv z(x=0)$ of the stellar centroid from the halo centre:
\begin{equation}\label{eq:rstar}
\frac{M(<r_*)}{r_*^2} \: \hat{r}_* = \vec{a}_5 \: \frac{\Delta G}{G_N^2},
\end{equation}
where $M(<r)$ is the enclosed dark matter plus gas mass. We now generalise this calculation to require equilibrium at an arbitrary point along the disk. The condition is
\begin{equation}\label{eq:resolve}
a_{5,z} \frac{\Delta G}{G_N} = a_{h,z} = -a_h \frac{z}{r},
\end{equation}
where $r = \sqrt{z^2 + x^2}$, $a_h$ is the magnitude of the acceleration due to the halo and subscript `$,z$' denotes $z$-projection. Assuming the halo is spherically-symmetric ($a_h = G_N M(<r)/r^2$) the disk must satisfy
\begin{equation}\label{eq:z1}
z(x) = -a_{5,z} \: \frac{\Delta G}{G_N^2} \frac{r^3}{M(<r)}.
\end{equation}
Thus negative $a_{5,z}$, as in Fig.~\ref{fig:cartoon}, implies positive $z$. The amplitude of this effect depends on the same function of the external fifth-force field, scalar coupling and total density profile as the offset $\vec{r}_*$ in Eq.~\ref{eq:rstar}, viz $a_5 \Delta G / M(<r)$, but is now a function of position along the disk. For realistic halo density profiles which fall with $r$, $z$ is an increasing function of $x$ so that the disk acquires a U shape bending away from $\vec{a}_{5,z}$.

We now assume that the gravitational restoring force is dominated by the dark matter, so that $M \simeq M_\text{halo}$, and that the warp is small so that $z(x) \ll x$. The first assumption is justified in Appendix~\ref{appendix} and the second is justified post-facto from the warp strengths resulting from reasonable fifth-force models (Sec.~\ref{sec:correlations}). This simplifies the above equation to\footnote{Note that in this approximation $z=0$ at $x=0$, so that Eq.~\ref{eq:rstar} is not reproduced. An overall offset is not of interest here but only the variation of $z$ with $x$; in practice we will redefine the $z$-axis anyway, so that the mean position of the warped disk is at $z'=0$. (This also increases independence from D18.)}
\begin{equation}\label{eq:w1_curve}
z(x) = -a_{5,z} \: \frac{\Delta G}{G_N^2} \frac{|x|^3}{M_\text{halo}(<x)}.
\end{equation}

For given $a_5$, the strength of the warp is maximised when the disk falls face-on in the external fifth-force field. Conversely, if the disk falls edge-on no warping in the direction of the disk normal is expected, although asymmetries will develop in the plane of the disk and in its kinematics~\cite{Jain_Vanderplas}.

\subsection{Summarising the warp curve}
\label{sec:statistic}

To simplify our inference and minimise the impact of noise we compress the warp curve into a scalar summary statistic. Our choice is motivated by the desire to maximise sensitivity to U-shaped warps, which are predicted by the fifth-force model and less commonly observed with large magnitude than S-shaped ones. Following~\cite{Secco}, we choose
\begin{equation} \label{eq:w1_integral}
w_1 \equiv \frac{1}{L_x^3} \int_{-L_x}^{L_x} |x| \: z(x) \: dx
\end{equation}
where $L_x$ is the distance along the major axis from the centre of the galaxy out to which the warp curve is calculated. The choice of $L_x$ is a trade-off between maximising the information included from the outer regions of the disk where the warp is most pronounced, on the one hand, and reducing the contamination from low signal to noise regions where sky fluctuations dominate, on the other. Our fiducial choice is $L_x = 3 R_\text{eff}$ (with $R_\text{eff}$ the half-light radius of the disk) although we check that qualitatively similar results follow from 2 or 4 $R_\text{eff}$. $w_1$ is unitless and zero for an S-shaped warp. The determination of $w_1$ in the observational data is described in Sec.~\ref{sec:observed_w1}; here we continue to focus on the model prediction.\footnote{The compression of the warp curve into $w_1$ necessarily throws away information, and may therefore be expected to weaken our inference. It is not clear however whether finer-grained features of the warp curve possess significant additional constraining power over the noise: we leave investigation of this issue to future work.}

\begin{figure}
  \centering
  \includegraphics[width=0.5\textwidth]{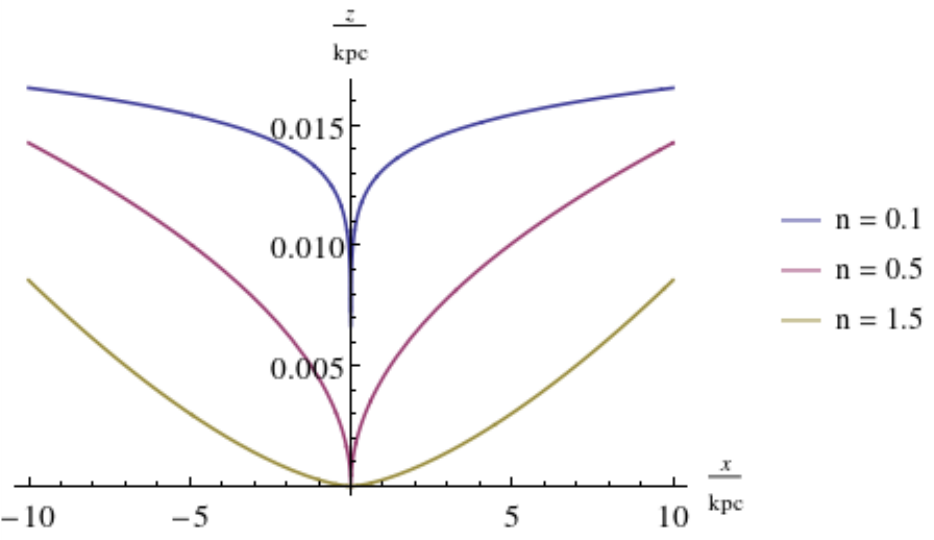}
  \caption{Expected warp curves deriving from a fifth force according to Eq.~\ref{eq:z}, as a function of the power-law slope $n$ of the halo density profile within the stellar disk. To evaluate $K$ we take typical illustrative values $r_s = 10$ kpc, $\rho_{rs} = 10^7 M_\odot$ kpc$^{-3}$, $a_{5,z} = 10^{-16}$ km/s and $\Delta G/G_N=1$. The disk's major axis lies along $x$, and $z$ is the orthogonal direction on the plane of the sky. Fifth-force warping is an $\mathcal{O}(10$ pc) effect: sensitivity to it requires a statistical sample of galaxies in a range of gravitational environments.}
  \label{fig:mathematica}
\end{figure}

To perform Bayesian inference we must forward-model the signal, which requires solving Eqs.~\ref{eq:w1_curve} and~\ref{eq:w1_integral} to find the expected $w_1$ for given fifth-force and galaxy parameters. As galaxies occupy the central regions of halos (typically within the scale radius), we begin by approximating the dark matter density over the extent of the galaxy by a single power-law in $r$.\footnote{We have checked that $r_s>3R_\text{eff}$ for the great majority of galaxies in most abundance matching realisations (see Sec.~\ref{sec:expected_w1}).} To allow for more general profiles than NFW (e.g. to take account of the effect of baryonic feedback), we keep the power-law index free and assign a specific value only when a numerical answer is required:
\begin{equation}\label{eq:n}
\rho(r) \simeq \rho(r_s)\:(r/r_s)^{-n}.
\end{equation}
We pivot the power law at the halo scale radius $r_s$ because this is typically well-measured in NFW fits to the halos produced in N-body simulations. These will be employed below to estimate halo properties, in conjunction with empirical models for the galaxy--halo connection. This gives an enclosed mass
\begin{equation}\label{eq:M}
M(<R) = \frac{4 \pi \rho_{rs}}{3-n} \: r_s^n \: R^{3-n},
\end{equation}
where $\rho_{rs} \equiv \rho(r_s)$. Assuming that the warp is small ($z \ll x$, justified in Sec.~\ref{sec:correlations}),
\begin{equation}\label{eq:R}
R = (x^2 + z^2)^{1/2} \simeq x +\mathcal{O}(z^2).
\end{equation}
Plugging Eqs.~\ref{eq:M} and~\ref{eq:R} into Eq.~\ref{eq:w1_curve} we find the warp curve
\begin{equation}\label{eq:z}
z(x) \simeq K \: |x|^n,
\end{equation}
where
\begin{equation}
K \equiv -\frac{a_{5,z} \Delta G}{G_N^2} \: \frac{3-n}{4 \pi \rho_{rs} r_s^n}.
\end{equation}
We show example warp curves for three different values of $n$ in Fig.~\ref{fig:mathematica}.

To facilitate comparison with the observations we now define a new variable $z'$ with a mean of $0$ across the disk out to $L_x \equiv 3 R_\text{eff}$:
\begin{equation}
\langle z \rangle = \frac{1}{L_x} \int^{L_x}_0 z(x) \: dx = \frac{K}{n+1} \: L_x^n,
\end{equation}
so that
\begin{equation}
z'(x) \equiv z(x) - \langle z \rangle = K\left(|x|^n - \frac{L_x^n}{n+1}\right).
\end{equation}
Finally we evaluate the integral for $w_1$ in these transformed coordinates:
\begin{align} \label{eq:w1}
w_1 &= \frac{2}{L_x^3} \int^{L_x}_0 z'(x) \: x \: dx = \frac{2 K}{L_x^3} \left(\frac{L_x^{n+2}}{n+2} - \frac{L_x^{n+2}}{2(n+1)}\right)\\ \nonumber
&= K \: \frac{n}{(n+1)(n+2)} \: L_x^{n-1}.
\end{align}
where we have used the symmetry of the warp about the midpoint of the disk to halve the range of the integral. Substituting for $L_x$ and $K$ yields
\begin{equation}\label{eq:w1_final}
w_1 = - \frac{n(3-n)}{(n+1)(n+2)} \: a_{5,z} \: \frac{\Delta G}{G_N^2} \: \frac{1}{4 \pi \rho_{rs}} \: \frac{(3 R_\text{eff}/r_s)^n}{3R_\text{eff}}.
\end{equation}

\section{Observational data}
\label{sec:data}

We use the \textit{Nasa Sloan Atlas} (NSA),\footnote{\url{http://www.nsatlas.org/}} a compilation of UV, optical and near-IR data for galaxies within $\sim250$ Mpc. To parallel D18, we will be interested in fifth-force ranges up to $50$ Mpc, which, given that our screening maps extend only to $\sim200$ Mpc~\cite{Desmond}, forces us to restrict our observational dataset to $D \lesssim 100$ Mpc. This is because we need to know the screening properties not only of the test galaxies, but also of the objects sourcing the fifth forces felt by those galaxies, as screened objects do not contribute to $\vec{a}_5$. We also cut at $M_* > 10^9 M_\odot$, necessary for precise characterisation of the halos of the test galaxies as required for evaluating Eq~\ref{eq:w1_final}. We use the $r$-band images to measure the warp (Sec.~\ref{sec:observed_w1}), and record also galaxy stellar mass, $M_*$, apparent minor-to-major axis ratio, $b/a$, angular half-light radius, \texttt{SERSIC\_TH50} and angle of the disk's major axis relative to North, \texttt{SERSIC\_PHI}, which we will use in modelling the warps (Sec.~\ref{sec:expected_w1}). We show three example raw images in Fig.~\ref{fig:images}.

In Sec.~\ref{sec:signal} we modelled test galaxies as thin disks oriented exactly edge-on. Although it is possible to restrict our observational sample to such disks ($b/a=0.15$), this would drastically reduce our statistics. A puffy disk that is oriented edge-on may continue to be analysed with the method of Sec.~\ref{sec:derivation}, provided that $z(x)$ is taken to be the centre of mass across the slice of the disk at $x$ and that tides across the galaxy are not significant. As most of our galaxies are isolated this should be a reasonable approximation. For a galaxy not oriented fully edge-on it is more difficult to predict exactly how the warp should appear. As a compromise between statistics and possible systematic error we cut at $b/a=0.5$. Although we cannot gauge the uncertainty in the expected $w_1$ ensuing from a fifth force for galaxies of more complex geometry, we can to first order model the change in contribution from other physics by introducing a dependence on $b/a$ of the width of the non-fifth-force part of the likelihood function. This is described in Sec.~\ref{sec:likelihood}. With these cuts our final sample contains $4,206$ galaxies at a mean distance of 74 Mpc.

\section{Method}
\label{sec:method}

Our overall method is similar to D18 in which further details may be found. We forward-model $w_1$ as a function of the global parameters $\Delta G/G_N$ and $\lambda_C$ and the galaxy-specific parameters $R_\text{eff}, r_s, \rho_{rs}, \Phi$ and $\vec{a}_5$. We specify probability distributions for the galaxy-specific parameters to create a fifth-force likelihood function $\mathcal{L}_5(w_1|\Delta G/G_N, \lambda_C)$ for each NSA galaxy, convolve it with a Gaussian that models the effect of non-fifth-force physics and constrain $\{\lambda_C, \Delta G/G_N\}$ by Markov Chain Monte Carlo (MCMC). We begin with our measurements of $w_1$ from the NSA sample before detailing our forward model.

\subsection{Measuring $w_1$}
\label{sec:observed_w1}

We begin by rotating each image through \texttt{SERSIC\_PHI} so that the galaxy's major axis is along the direction we designate as $x$. Taking the perpendicular axis on the plane of the sky to be $z$, we set the origin of our coordinate system at the centroid of the galaxy's $r$-band emission and use units of pixel size. We scan in the $x$ direction through all pixels between $\pm 3 R_\text{eff}$, and for each vertical slice calculate the intensity-weighted mean value of $z$, $\bar{z}(x)$, between $z = \pm \: L_z \equiv 3\:(b/a)\:R_\text{eff}$:
\begin{equation}
\bar{z}(x) = \frac{\sum_{z=-L_z}^{z=L_z} z \: I(x,z)}{\sum_{z=-L_z}^{z=L_z} \: I(x,z)},
\end{equation}
where the sum is over pixels and $I(x,z)$ is the measured flux in pixel ($x,z$). For the observational data $\bar{z}$ replaces $z$ in Eqs.~\ref{eq:w1_integral} and~\ref{eq:w1_final} as a measure of the average position of the disk at given $x$. The choice of $L_z$ is intended to roughly cover the extent of the disk along its minor axis; we have checked that our conclusions are qualitatively unaffected by changing $L_z$ within $2-4$ $(b/a) \: R_\text{eff}$.

\begin{figure*}
  \subfigure[J115012.10+065956.9, $\;b/a=0.15$]
  {
    \includegraphics[width=0.317\textwidth]{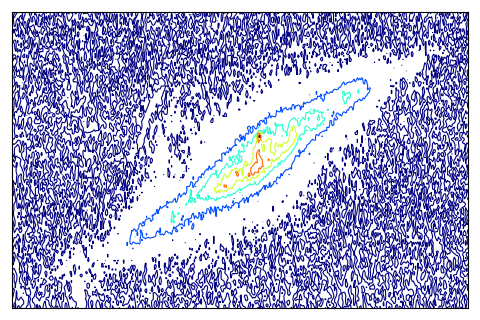}
    \label{fig:image1}
  }
  \subfigure[J003938.31+143951.2, $\;b/a=0.27$]
  {
    \includegraphics[width=0.317\textwidth]{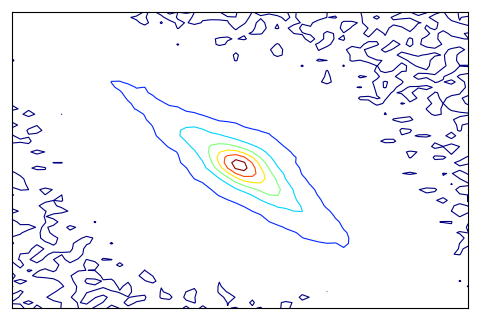}
    \label{fig:image2}
  }
  \subfigure[J123643.71-030114.3, $\;b/a=0.43$]
  {
    \includegraphics[width=0.317\textwidth]{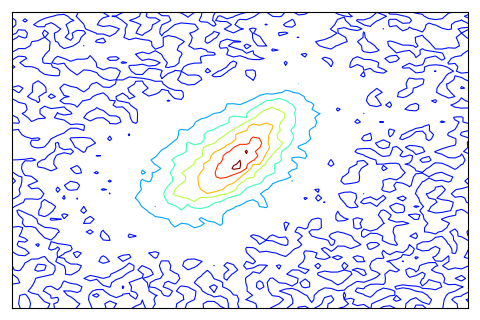}
    \label{fig:image3}
  }
  \caption{Example $r$-band images of galaxies from the \textit{Nasa Sloan Atlas}, labelled by IAU designation and apparent axis ratio $b/a$.}
  \label{fig:images}
\end{figure*}

\begin{figure*}
  \subfigure[$w_1 = 1.7\times10^{-3}$]
  {
    \includegraphics[width=0.317\textwidth]{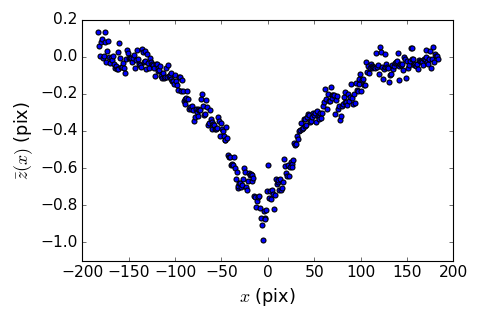}
    \label{fig:warp1}
  }
  \subfigure[$w_1 = -1.1\times10^{-4}$]
  {
    \includegraphics[width=0.317\textwidth]{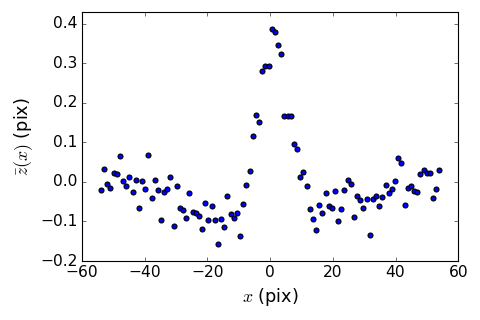}
    \label{fig:warp2}
  }
  \subfigure[$w_1 = 7.4\times10^{-4}$]
  {
    \includegraphics[width=0.317\textwidth]{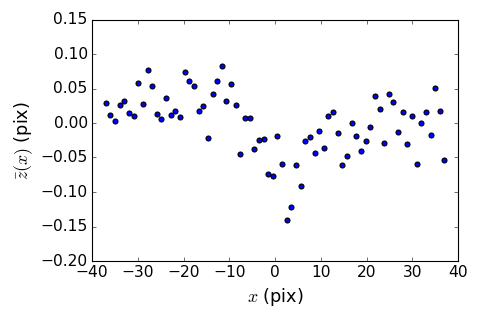}
    \label{fig:warp3}
  }
  \caption{Warp curves out to $\pm 3 R_\text{eff}$ derived from the images in Fig.~\ref{fig:images}, with corresponding $w_1$ calculated from Eq.~\ref{eq:w1_obs}.}
  \label{fig:warps}
\end{figure*}

The warp curves $\bar{z}(x)$ of the galaxies in Fig.~\ref{fig:images} are shown in Fig.~\ref{fig:warps}. These show that there is some offset between the origin of the coordinate system and the centre of the disk; to remove the bias this would induce in $w_1$ we redefine the $z$-axis so that the centre of the warp curve is at $z'=0$:
\begin{eqnarray}
\begin{aligned}
&\langle z \rangle = \frac{1}{6 R_\text{eff}} \sum_{x=-3R_\text{eff}}^{x=3R_\text{eff}} \bar{z}(x),\\
&\bar{z}'(x) \equiv \bar{z}(x) - \langle z \rangle.
\end{aligned}
\end{eqnarray}
Finally we evaluate Eq.~\ref{eq:w1_integral} as
\begin{equation}\label{eq:w1_obs}
w_1 = \frac{1}{(3R_\text{eff})^3} \sum_{x=-3R_\text{eff}}^{x=3R_\text{eff}} |x| \: \bar{z}'(x).
\end{equation}

\subsection{Forward-modelling $w_1$}
\label{sec:expected_w1}

As $w_1$ vanishes in overall-screened galaxies and is otherwise proportional to the fifth-force field $\vec{a}_5$, the first step to forward-modelling it on a galaxy-by-galaxy basis is to precisely characterise galaxies' gravitational environments. To do so we utilise the maps of~\cite{Desmond}, where both $\Phi_\text{ex}$ and $\vec{a}_5$ are calculated as a function of fifth-force range $\lambda_C$ at every point in space to a distance $\sim200$ Mpc. \cite{Desmond} combines the contributions from mass associated with galaxies in the 2M++ all-sky redshift survey \cite{2M++} (linked to halos by abundance matching), resolved halos in an N-body simulation hosting galaxies too faint to be observed (the effect of which is correlated with observables to model point-by-point in the real Universe), and mass distributed in linear and quasi-linear modes of the density field not captured by the halo model. The latter derives from the BORG algorithm~\citep{Jasche_10, Jasche_15, Jasche_Wandelt_12, Jasche_Wandelt_13, Lavaux, BORG_PM}, which propagates information from the 2M++ galaxy number density field to the current total density field and its initial conditions by means of a Bayesian likelihood framework, assuming best-fit Planck $\Lambda$CDM cosmology and marginalising over a bias model. The parameters forming the inputs to this calculation are drawn from probability distributions describing their possible values, so that a probability distribution is derived for $\Phi_\text{ex}$ and $\vec{a}_5$ at the position of each test galaxy rather than a single value. Further details on our deployment of these maps may be found in D18 secs. IV A-C, and the distributions of the input parameters themselves are summarised in table 1 (rows 1-4) of that paper. We project $\vec{a}_5$ onto the normal of each disk (on the plane of the sky) using:
\begin{equation}
a_{5,z} = a_{5,\delta} \sin(\theta) - a_{5,\alpha} \cos(\theta),
\end{equation}
where $a_{5,\alpha}$ and $a_{5,\delta}$ are the right ascension (RA) and declination (DEC) components of $\vec{a}_5$ at the test point and $\theta=$ \texttt{SERSIC\_PHI} is the angle of the disk's major axis East of North on the sky.

The total screening proxy $\Phi$ is the sum of $\Phi_\text{ex}$ and an internal contribution $\Phi_\text{in}$ due to the galaxy's own mass. We take $\Phi_\text{in} = V_\text{disp}^2$, with $V_\text{disp}$ as recorded in the NSA, which we assign a $10\%$ uncertainty as in D18 (table 1 row 5). We draw $N_\text{MC}=1000$ Monte Carlo realisations of the model, each independently sampling the $\Phi_\text{in}$, $\Phi_\text{ex}$ and $\vec{a}_5$ distributions separately for each test galaxy. For a given $\lambda_C$ this gives a probability $f \equiv N(|\Phi| < |\Phi_c|)/N_\text{MC}$ for the galaxy to be unscreened, where $|\Phi_c| = 1.5 \times 10^{-4} \: (\lambda_C/32\:\text{Mpc})^2$ is the screening threshold in $\Phi$ for Hu-Sawicki $f(R)$~\citep{Hu_Sawicki}.

Next we estimate $r_s$, $\rho_{rs}$ and $R_\text{eff}$ for each test galaxy. We begin by performing abundance matching (AM;~\cite{Kravtsov,Conroy,Reddick}) to assign a stellar mass to each halo in the \textsc{DarkSky-400} simulation \cite{DarkSky}, post-processed with the \textsc{Rockstar} halo finder \cite{Rockstar}. We use the stellar mass function of~\cite{Bernardi_SMF}, which has been shown to match the stellar mass determinations of the NSA from \textsc{kcorrect}, and the specific AM parametrisation of~\cite{Lehmann} which is known to accurately reproduce galaxy clustering as a function of both stellar mass and $r$-band luminosity. We use the best-fit parameters $\alpha = 0.6$ and $\sigma_\text{AM} = 0.16$ for the AM proxy and universal Gaussian scatter in stellar mass at fixed halo mass, respectively. We then associate each test galaxy with the object in the abundance-matched catalogue closest to it in $M_*$, taking $r_s$ from the \textsc{Rockstar} output and calculating $\rho_{rs}$ assuming an NFW profile. As AM is inherently stochastic when $\sigma_\text{AM}>0$, we generate $200$ mock catalogues in this way and randomly select one for each Monte Carlo realisation of our model. $R_\text{eff}$ is derived by multiplying the angular major-axis half-light radius of a 2D S\'{e}rsic fit to the light profile (\texttt{SERSIC\_TH50} in the NSA) by the angular diameter distance to the galaxy (assuming a $\Lambda$CDM cosmology $h=0.7$, $\Omega_\Lambda = 0.7$, $\Omega_m = 0.3$, $\Omega_k = 0$) and circularising with a further factor of $(b/a)^{1/2}$.

We now possess all the information necessary to forward-model $w_1$ by Eq.~\ref{eq:w1_final} for each NSA galaxy as a function of $\Delta G/G_N$ and $\lambda_C$. We take a fiducial value for $n$ (Eq.~\ref{eq:n}) of $0.5$, corresponding to a central dark matter density slope that has been flattened somewhat from a primordial NFW profile by baryonic feedback. This is reasonable in light of both observations and simulations (e.g.~\citep{cusp-core, Rodrigues}), although we check that varying $n$ in the range $0.1-1.5$ has at most a factor of 4 effect on maximum-likelihood $\Delta G/G_N$ values.

\subsection{Likelihood model and inference method}
\label{sec:likelihood}

Each Monte Carlo realisation of our model produces a possible value of $w_1$ for given $\{\lambda_C, \Delta G/G_N\}$, so that the full set of realisations specifies the likelihood. Rather than assume a specific form for this we approximate it empirically as a histogram by dividing the $N_\text{MC}$ $w_1$ values into $N_\text{bins}=10$ equal-width bins between the smallest and largest values $w_{1(0)}$ and $w_{1(1)}$:
\begin{align} \label{eq:L_F5}
\mathcal{L}_{5,i}(w_1|&\Delta G, \lambda_C) = (1-f) \: \delta(w_1) \\ \nonumber
&+ f \: \Sigma_{j=0}^{N_\text{bins}-1} \: P_j \: \delta(w_1 - \Delta G \: w_j)
\end{align}
where $i$ is the galaxy index, $\Delta w_1 \equiv (w_{1(1)} - w_{1(0)})/N_\text{bins}$ is the bin width, $w_j \equiv w_{1(0)} + \: (j+1/2) \:\Delta w_1$ is the position of the jth bin centre and $P_j \equiv N_j/N_\text{MC}$ is the fraction of model realisations falling in bin $j$. The first term gives the screened contribution to $\mathcal{L}_5$ and the second term the unscreened contribution.

The full likelihood function however ought to contain an additional component that permits $w_1$ to be non-zero even in the absence of a fifth force: this describes measurement error in $w_1$ as well as a contribution from any other type of physics. We will see in Sec.~\ref{sec:correlations} that by Eq.~\ref{eq:w1_final} our best-fit fifth-force models produce $w_1$ values on average two orders of magnitude smaller than those observed, so that this additional likelihood component is in fact required to provide the bulk of the signal. For want of additional information we take this component to be Gaussian and choose its width such that it accounts on average for the entire signal. This conservative choice guarantees that to first order the entire dataset can be accounted for without fifth forces.

\begin{figure}
  \centering
  \includegraphics[width=0.5\textwidth]{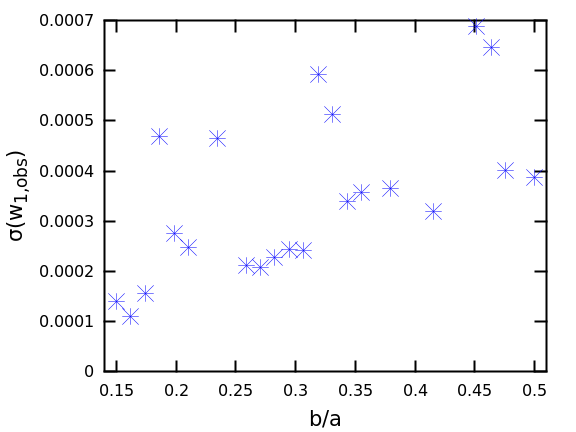}
  \caption{Dispersion in observed $w_1$ plotted against apparent galaxy axis ratio $b/a$ in the NSA sample. We take 30 bins between $b/a=0.15$ and $0.5$. The strong trend and large scatter indicate that this correlation must be taken into account in the non-fifth-force part of the likelihood model.}
  \label{fig:ba}
\end{figure}

The simplest choice would be to take this width to be the same for all galaxies and equal to the overall standard deviation in observed $w_1$, $\sigma(w_{1,\text{obs}})$. However, this would bias our inference if the dispersion correlates significantly with galaxy properties. We find $\sigma(w_{1,\text{obs}})$ to correlate most strongly with the apparent minor-to-major axis ratio of the disk $b/a$ (Fig.~\ref{fig:ba}), describing the variation in measured warp strength when the disk is inclined from edge-on or has significant thickness. Neglecting this correlation would cause $\sigma(w_{1,\text{obs}})$ to be overestimated at small $b/a$ and underestimated at large $b/a$, skewing the part of the measured $w_1$ for which the fifth-force component of the likelihood is required to account. We therefore make the width of the Gaussian noise component of the likelihood a function of $b/a$ by setting it equal to $\sigma(w_{1,\text{obs}})$ separately in $30$ bins of $b/a$ between the minimum and maximum values $0.15$ and $0.5$. This ensures that for any $b/a$ the data can be fully accounted for to good accuracy regardless of the behaviour of the fifth force, provided only that this force provides a subdominant contribution to $w_1$. To leading order, then, $w_1$ is purely noise.

Convolving $\mathcal{L}_{5,i}$ in Eq.~\ref{eq:L_F5} with this Gaussian (of width $\sigma_i$ for galaxy $i$) yields the total likelihood function
\begin{align} \label{eq:Ltot}
&\mathcal{L}_i(w_1|\Delta G, \lambda_C) = (1-f) \: \frac{\exp\{-w_1^2/2\sigma_i^2\}}{\sqrt{2 \pi} \: \sigma_i} \\ \nonumber
& + f \: \Sigma_{j=0}^{N_\text{bins}} \: P_j \: \frac{\exp\{{-(w_1 - \Delta G \: w_j)^2/2\sigma_i^2}\}}{\sqrt{2 \pi} \: \sigma_i}.
\end{align}

We find practically identical constraints on $\{\Delta G/G_N, \lambda_C\}$ if $\sigma_i$ is parametrised as a function of $b/a$ and fitted along with the fifth-force parameters. Our noise model is formally identical to that of D18 (secs. IV E-F), except that it depends on $b/a$ rather than signal to noise ratio $s$.

\vspace{5mm}

\noindent We show three separate principal analyses. For the first two we perform 1D inference for $\Delta G/G_N$, separately for $20$ logarithmically-uniformly spaced values of $\lambda_C$ in the range $0.4-50$ Mpc. In the first we consider a screened fifth-force model as described above. In the second we switch screening off so that the unscreened fraction $f$ for each test galaxy is $1$ and all mass within $\lambda_C$ contributes to $\vec{a}_5$. (For this to lead to warping the fifth force must couple to the dark matter but not the stars, which is likely contrived; we show this merely as a foil for the screening case.) For the third analysis we again include screening and perform a full 2D inference of both $\Delta G/G_N$ and $\lambda_C$.

We check that our analysis is converged with number of Monte Carlo realisations of the model, that the AM galaxy--halo connection and smooth density field from BORG are thoroughly sampled, that our MCMC is converged with the number of steps, and that our principal results are insensitive to reasonable variations in $M(<r)$ and the assumed uncertainties in galaxy and halo properties. Further validation is documented in Sec.~\ref{sec:valid}.

\section{Results}
\label{sec:results}

\subsection{Comparison of observed and predicted signal}
\label{sec:correlations}

\begin{figure*}
  \subfigure[]
  {
    \includegraphics[width=0.48\textwidth]{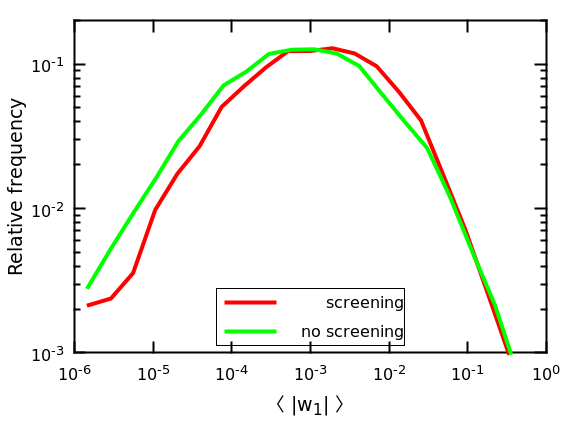}
    \label{fig:1a}
  }
  \subfigure[]
  {
    \includegraphics[width=0.48\textwidth]{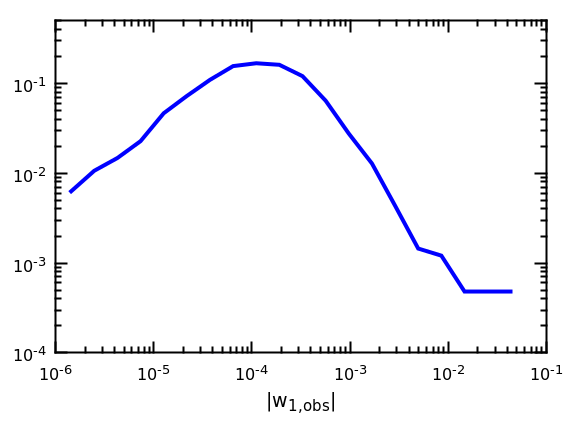}
    \label{fig:1b}
  }
  \subfigure[]
  {
    \includegraphics[width=0.48\textwidth]{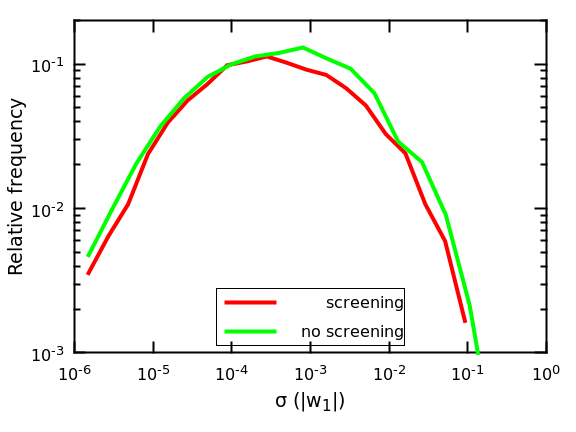}
    \label{fig:1c}
  }
  \subfigure[]
  {
    \includegraphics[width=0.48\textwidth]{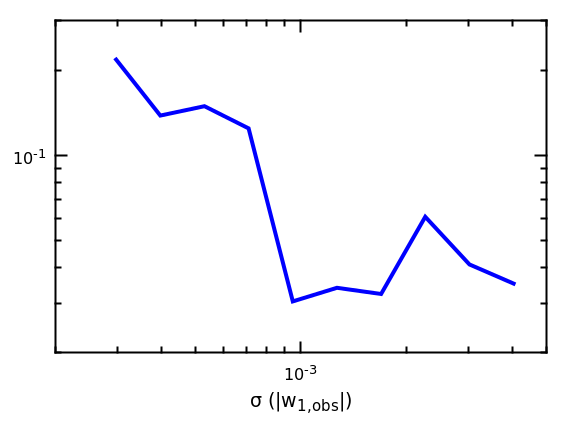}
    \label{fig:1d}
  }
  \caption{Histograms of the expectation (upper) and uncertainty (lower) of the fifth-force predicted (left) and observed (right) signal $w_1$ over all galaxies in our dataset. $\langle |w_1| \rangle$ is the modal average of 1000 Monte Carlo realisations of the model with $\Delta G/G_N = 1$, $\lambda_C = 5$ Mpc, given 10 bins of $w_1$ between its maximum and minimum values. The uncertainty is the minimal width enclosing $68\%$ of the model realisations, which is the width of the fifth-force part of the likelihood function. For the observed signal the uncertainty is the standard deviation of $w_1$ conditioned on apparent axis ratio $b/a$ (see Sec.~\ref{sec:likelihood}). This is the width of the non-fifth-force part of the likelihood function.}
  \label{fig:1}
\end{figure*}

In Fig.~\ref{fig:1} we compare the measured $|w_1|$ values and their scatter to those predicted by a model with $\lambda_C = 5$ Mpc, $\Delta G/G_N = 1$, both with and without screening. The expectation $\langle |w_1| \rangle$ is the mode of the $N_\text{MC}$ realisations of the model (unscreened component only), which is the location of the bin with maximum $P_j$ (Eq.~\ref{eq:Ltot}). $\sigma(|w_1|)$ is the minimal width enclosing $68\%$ of the realisations and $\sigma(|w_{1,\text{obs}}|)$ is the $b/a$-dependent scatter described in Sec.~\ref{sec:likelihood}. We see that the values predicted by this model are typically a factor of a few larger than the observed ones, indicating that if the predicted and observed warps for individual galaxies were aligned then a fifth force of this strength could account for the entire signal. By contrast, for the maximum-likelihood $\Delta G/G_N$ values of D18 ($\sim0.02$, which we will also find to be of most interest here), $\langle |w_1| \rangle$ is over an order of magnitude smaller than $|w_{1,\text{obs}}|$ on average, indicating that the better part of each warp derives from non-fifth-force effects. In this case our approximation of Sec.~\ref{sec:likelihood} is valid and the fifth-force contribution to the likelihood can be considered a small perturbation to the noise. That the $\langle |w_1| \rangle$ and $\sigma(w_1)$ distributions are similar between the screening and no screening runs at this $\lambda_C$ indicates that most of the mass sourcing $\vec{a}_5$ is unscreened in either case.

\begin{figure*}
  \subfigure[Predicted $w_1-\Phi$]
  {
    \includegraphics[width=0.48\textwidth]{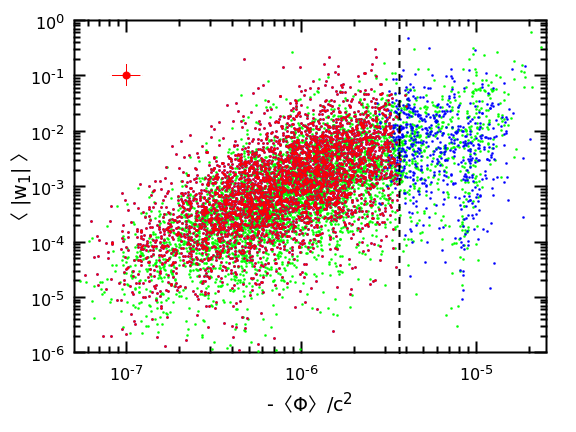}
    \label{fig:2a}
  }
  \subfigure[Predicted $w_1-a_5$]
  {
    \includegraphics[width=0.48\textwidth]{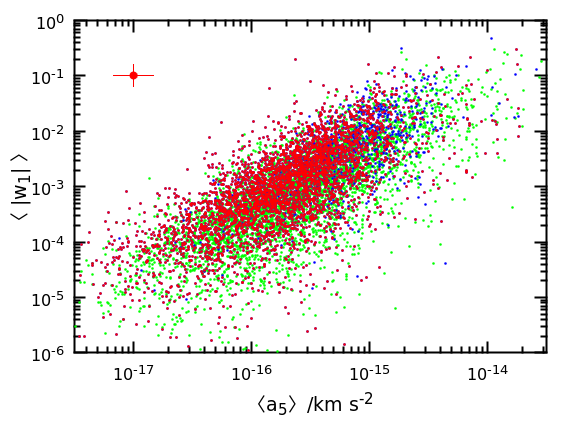}
    \label{fig:2b}
  }
  \subfigure[Observed $w_1-\Phi$]
  {
    \includegraphics[width=0.48\textwidth]{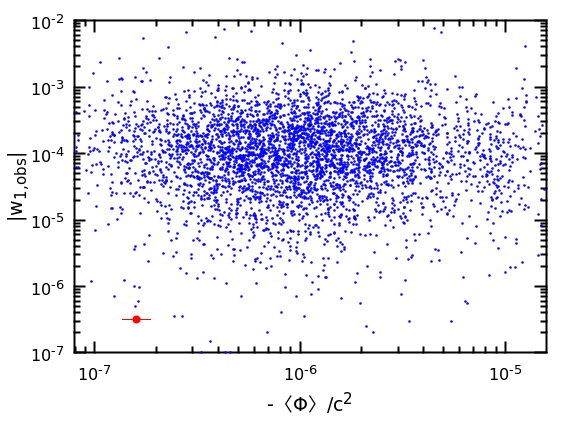}
    \label{fig:2c}
  }
  \subfigure[Observed $w_1-a_5$]
  {
    \includegraphics[width=0.48\textwidth]{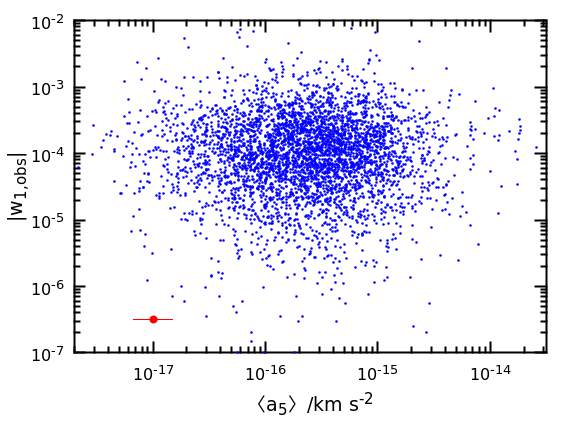}
    \label{fig:2d}
  }
  \caption{Correlations of predicted (upper) and observed (lower) signals with Newtonian potential $\Phi$ (left) and fifth-force acceleration $a_5$ (right). The calculation of the prediction is as Fig.~\ref{fig:1} for the fiducial case $\lambda_C = 5$ Mpc, $\Delta G/G_N=1$, with $\langle \Phi \rangle$ and $\langle a_5 \rangle$ the modal averages over the 1000 model realisations. In the top row, the green points show the case in which screening is switched off, the red points for when it is included and the blue points as the red except without the $w_1$ values of screened galaxies set to 0. The black dashed line shows the screening threshold $|\Phi_c|$ for this model. The median sizes of the errorbars in the four directions (minimal 68\% bounds for the prediction, Gaussian $1\sigma$ for the observations) are shown by the red crosses, although we suppress the vertical uncertainty in the bottom panels as it is simply the scatter in the points.}
  \label{fig:2}  
\end{figure*}

\begin{figure}
  \centering
  \includegraphics[width=0.5\textwidth]{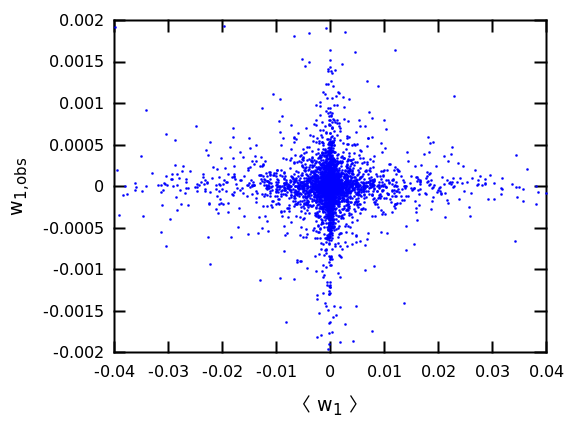}
  \caption{Correlation of predicted and observed $w_1$ for the model of Figs.~\ref{fig:1} \& \ref{fig:2}.}
  \label{fig:3}
\end{figure}

In Fig.~\ref{fig:2} we show the correlations of the predicted and observed $|w_1|$ values with the Newtonian potential $|\Phi|$ and fifth-force acceleration $a_5$. As in D18, the predicted signal rises with $|\Phi|$ and $a_5$, cuts off in the screened case at $|\Phi| > |\Phi_c|$ when the test galaxies become screened (red), and is otherwise slightly reduced by screening of the source masses (blue). The Spearman's, $\rho_\text{s}$, and Pearson's, $\rho_\text{p}$, coefficients for the blue points are 0.60 and 0.57 respectively for the correlation with $\Phi$, and 0.74 for the correlation with $a_5$. The green points are for the model in which screening is entirely switched off. The data on the other hand show no such correlations ($\rho_\text{s}$, $\rho_\text{p} = 0.0$; bottom panels), indicating again that a fifth force can account for at most a small part of the signal. This requires $\Delta G/G_N \ll 1$. Fig.~\ref{fig:3} plots the predicted and observed $w_1$ directly against one another.

\subsection{Constraints on $\lambda_C$ and $\Delta G/G_N$}
\label{sec:constraints}

\begin{figure}
  \centering
  \includegraphics[width=0.5\textwidth]{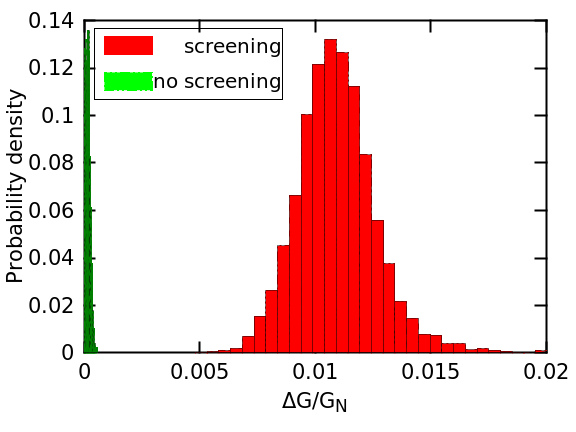}
  \caption{Posterior of $\Delta G/G_N$ for $\lambda_C = 1.8$ Mpc, both with and without screening. As explained in Sec.~\ref{sec:valid}, we take the screened model of this Compton wavelength to provide the best overall solution to our inference.}
  \label{fig:dG3}
\end{figure}
\begin{figure*}
  \subfigure[$\lambda_C = 500$ kpc]
  {
    \includegraphics[width=0.3\textwidth]{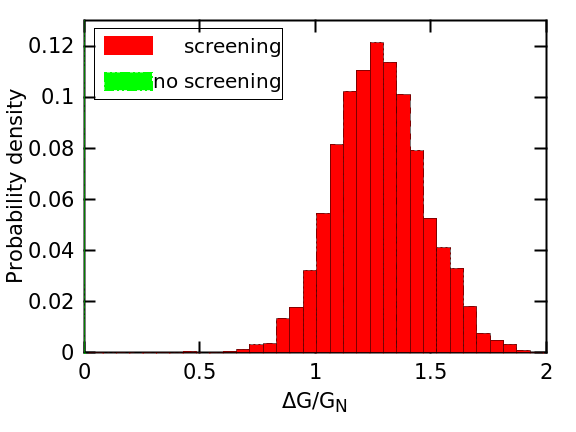}
    \label{fig:comp_a}
  }
  \subfigure[$\lambda_C = 5$ Mpc]
  {
    \includegraphics[width=0.3\textwidth]{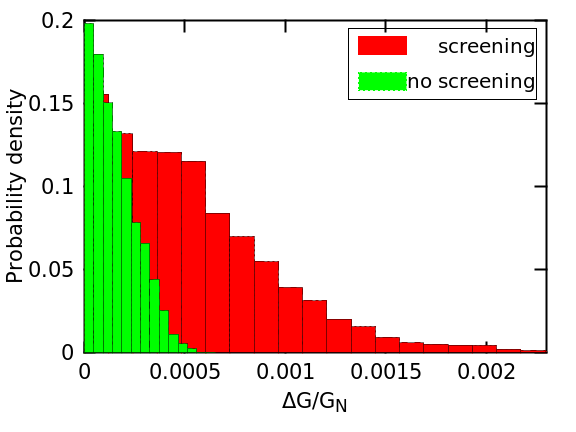}
    \label{fig:comp_b}
  }
  \subfigure[$\lambda_C = 50$ Mpc]
  {
    \includegraphics[width=0.3\textwidth]{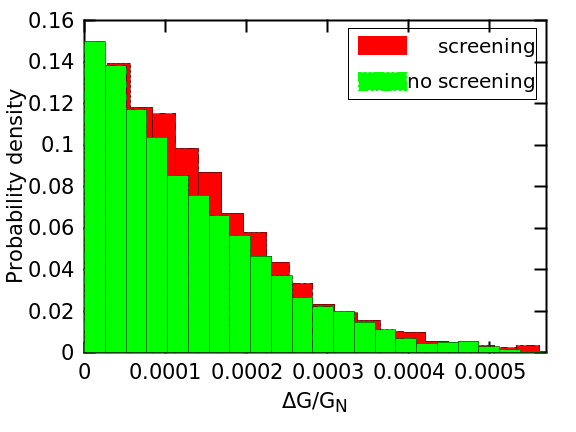}
    \label{fig:comp_c}
  }
  \caption{Posteriors on $\Delta G/G_N$ for three values of $\lambda_C$, as indicated in the captions, with and without screening. In Fig.~\ref{fig:comp_a} the posterior for the model without screening is at very small $\Delta G$.}
  \label{fig:comp}
\end{figure*}

We now employ the likelihood formalism of Sec.~\ref{sec:method} to constrain $\lambda_C$ and $\Delta G/G_N$. To begin, we fix $\lambda_C$ to one of its 20 values between $400$ kpc and $50$ Mpc and show in Figs.~\ref{fig:dG3} and~\ref{fig:comp} a range of posteriors for $\Delta G/G_N$ for models both with (red) and without (green) screening. In cases where the posteriors are peaked at $\Delta G/G_N=0$ the model without screening is more tightly constrained because screened galaxies are predicted to have $w_1=0$ and hence do not contribute to the constraint.

\begin{figure*}
  \subfigure[]
  {
    \includegraphics[width=0.48\textwidth]{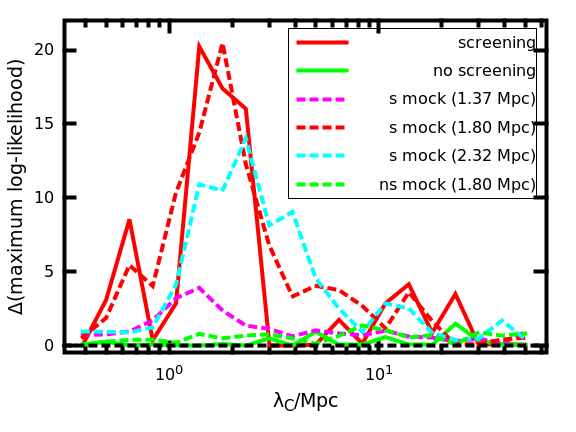}
    \label{fig:like_like}
  }
  \subfigure[]
  {
    \includegraphics[width=0.48\textwidth]{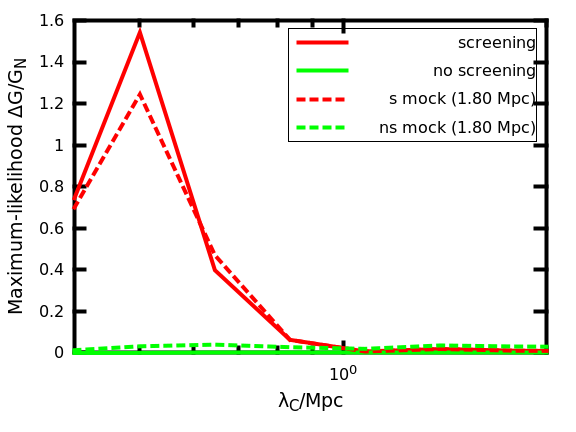}
    \label{fig:like_dg}
  }
  \caption{Fig.~\ref{fig:like_like} shows the maximum log-likelihood increase $\Delta\log(\mathcal{L})$ over the GR case $\Delta G=0$, for any $\Delta G/G_N$, separately for 20 $\lambda_C$ values in the range $0.4-50$ Mpc (solid lines). We show results both with and without screening. The model without screening (green) never achieves $\Delta\log(\mathcal{L})>2$, indicating both that this model cannot improve upon GR and that the noise in our experiment is under control. The model with screening on the other hand (red) achieves $\Delta\log(\mathcal{L}) > 15$ for $\lambda_C = 1.3-2.3$ Mpc. The dashed lines show the average results when the models are applied to 10 mock datasets with signals injected by hand at $\lambda_C = 1.37, 1.80$ or $2.32$ Mpc, as indicated in the legend, and the corresponding maximum-likelihood $\Delta G/G_N$ values 0.019, 0.011 or 0.0085. The magenta, red and cyan dashed lines are reconstructed by the model with screening (``s''), and the green dashed line by the model without (``ns''). That the red dashed and solid lines are similar over the entire range of $\lambda_C$ indicates that mock data with $\lambda_C=1.8$ Mpc and $\Delta G/G_N = 0.011$ is practically indistinguishable from the real data as seen by our inference framework. The smaller peaks in $\Delta\log(\mathcal{L})$ away from the preferred $\lambda_C$, also approximately reproduced by the mock data, may indicate correlations between the $\vec{a}_5$ fields resulting from different fifth-force ranges. Fig.~\ref{fig:like_dg} shows the best-fit $\Delta G/G_N$ values corresponding to the points of largest $\Delta\log(\mathcal{L})$ at each $\lambda_C$. At larger $\lambda_C$, $\Delta G/G_N \simeq 0$. The dashed lines show the results from mock data with $\lambda_C=1.8$ Mpc and $\Delta G/G_N = 0.011$, which gives similar results.}
  \label{fig:like}
\end{figure*}

In Fig.~\ref{fig:like_like} we show the increase in maximum log-likelihood $\Delta\log(\mathcal{L})$ over the General Relativistic (GR) case $\Delta G=0$ achieved for any $\Delta G/G_N$ at each $\lambda_C$ (sold lines; the dashed lines will be explained in Sec.~\ref{sec:valid}). We see a clear peak at similar $\lambda_C$ to D18, $1.3$--$2.3$ Mpc. Further, the best-fit $\Delta G/G_N$ values are in good agreement: the three points with highest likelihood are $\{\lambda_C/\text{Mpc}, \: \Delta G/G_N, \: \Delta\log(\mathcal{L})\} = \{1.37, 0.019, 20\}, \{1.80, 0.011, 17\}$ and $\{2.32, 0.0085, 16\}$. This corresponds to a maximum AIC difference \cite{AIC} of $36$ for a difference of two degrees of freedom, and hence a probability ratio of the baseline GR model to the screened model of $e^{-18}$. The $\Delta G/G_N$ posteriors are $\sim7\sigma$ discrepant with $\Delta G=0$. These results are highly statistically significant even with a trial factor $\sim1/50$ to account for the look-elsewhere effect. 

The models without screening achieve only very small $\Delta\log(\mathcal{L})$ for any $\lambda_C$, indicating that the improvement in the goodness-of-fit depends on the partition of both test and source galaxies into screened and unscreened subsets, which is a function of both their internal mass distribution and their environment. In conjunction with the mock data tests presented in Sec.~\ref{sec:valid} we take this to be valuable evidence concerning the validity of the fifth-force interpretation. Other effects would not be expected to correlate $w_1$ with $\Phi$ and $\vec{a}_5$ as screening does.

The trend in best-fit $\Delta G/G_N$ with $\lambda_C$ is shown in Fig.~\ref{fig:like_dg}. Models with larger $\lambda_C$ tend to give larger $\vec{a}_5$ because they incorporate the contribution from mass out to a larger distance from a given test galaxy, and hence prefer smaller $\Delta G/G_N$ for fixed $w_{1,\text{obs}}$. At larger $\lambda_C$ values than shown, $\Delta G/G_N \sim 0$.

\begin{figure*}
  \subfigure[]
  {
    \includegraphics[width=0.317\textwidth]{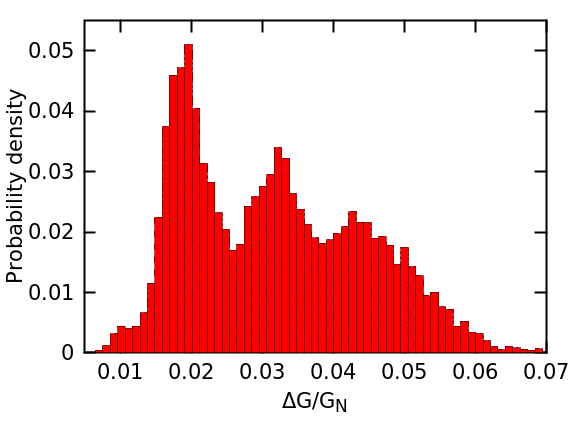}
    \label{fig:2D_dG}
  }
  \subfigure[]
  {
    \includegraphics[width=0.317\textwidth]{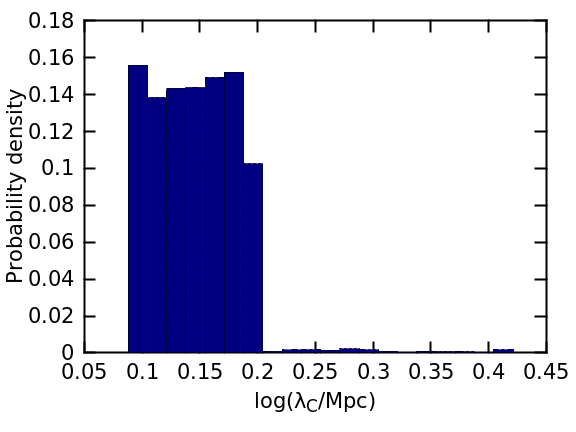}
    \label{fig:2D_lambda}
  }
  \subfigure[]
  {
    \includegraphics[width=0.317\textwidth]{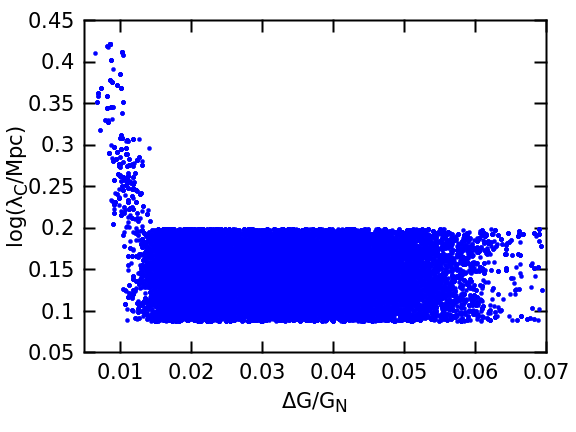}
    \label{fig:2D_comb}
  }
  \caption{Posteriors of $\Delta G/G_N$ and $\lambda_C$ from a 2D inference. Figs.~\ref{fig:2D_dG} and~\ref{fig:2D_lambda} show the marginalised constraints, while Fig.~\ref{fig:2D_comb} gives the full 2D posterior. The great majority of points in the chain lie in the $0.1 < \log_{10}(\lambda_C/\text{Mpc}) < 0.2$ bin.}
  \label{fig:2D}
\end{figure*}

Finally, Fig.~\ref{fig:2D} shows the constraints on $\Delta G/G_N$ and $\lambda_C$, and their covariance, for the full 2D inference in which both are free. Most of the probability density is located at the maximum-likelihood solution $\lambda_C = 1.37$ Mpc -- with $0.01 \lesssim \Delta G/G_N \lesssim 0.06$ -- although $\lambda_C$ up to $2.3$ Mpc is statistically marginally allowed. The trimodality in Fig.~\ref{fig:2D_dG} likely reflects the fact that in this case most of the model realisations fell into three non-adjacent bins of $w_1$ for many galaxies, each one preferring a slightly different value for $\Delta G/G_N$ to achieve the best fit to the data.

\subsection{Validation}
\label{sec:valid}

\subsubsection{Mock data with best-fit $\Delta G/G_N$}

We turn now to methods for validating our results. We begin by generating mock data with an imposed $\Delta G/G_N$ corresponding to one of the three maximum-likelihood solutions of Fig.~\ref{fig:like_like}. We do this by first calculating $\langle w_1 \rangle$ for each galaxy -- the mode of the model realisations if the galaxy is likely unscreened ($f>0.5$) and 0 otherwise -- and then scattering it by $\sigma_i$. This gives the mock data identical noise characteristics to the real data; were the fifth-force model correct the two datasets would also have identical correlations with gravitational environment. We then refit the model, first recalculating $\sigma_i$ by binning the mock $w_1$ values in $b/a$, as for the real data, and then repeating the MCMC to constrain $\Delta G/G_N$ at each $\lambda_C$. In particular, we record the maximum-likelihood $\Delta G/G_N$ and the increase in log-likelihood $\Delta\log(\mathcal{L})$ over $\Delta G=0$ that this achieves. We average these values over 10 mock datasets to sample the stochasticity induced by the noise and show the results as the dashed lines in Fig.~\ref{fig:like}: the magenta line uses mock data generated with $\{\lambda_C/\text{Mpc}, \Delta G/G_N\} = \{1.37, 0.019\}$, the red line $\{1.8, 0.011\}$, the cyan line $\{2.32, 0.0085\}$, and the green line as the red except refit by the model without screening.

Mock data with a signal at $\lambda_C = 1.37$ or $2.32$ Mpc cannot produce $\Delta\log(\mathcal{L})$ values as high as measured from the data, indicating that in these cases the noise provides too large a contribution to the overall likelihood relative to the signal. However, this is not true for data with a signal at $\lambda_C = 1.8$ Mpc, which gives an almost identical peak in $\Delta\log(\mathcal{L})$ at $1.3 \lesssim \lambda_C/\text{Mpc} \lesssim 2.3$ as the real data. Further, it roughly reproduces the smaller peaks seen at $\lambda_C \simeq 0.65$ and $14$ Mpc. This suggests that these peaks may not be noise in our experiment (which appears from the green lines in Fig.~\ref{fig:like_like} to be at the $\Delta\log(\mathcal{L}) \pm 1$ level) but rather arise from correlations at the test galaxy positions between the $\vec{a}_5$ and $\Phi$ fields resulting from the ``real'' fifth-force range $\sim1.8$ Mpc and those which would be generated from those other ranges. As these correlations must be dependent on the locations at which the fields are sampled, we would expect them to diminish as the number of galaxies in the sample (probing different gravitational environments) was increased. The $2.5$ times larger sample size in D18 than here may explain why subsidiary peaks in $\Delta\log(\mathcal{L})$ were not seen in figure 8a of that work. The use of a vector rather than scalar observable may also make it rarer for correlations between $\vec{a}_5(\lambda_C)$ and $\Phi(\lambda_C)$ at different $\lambda_C$ to propagate significantly into $\Delta\log(\mathcal{L})$.

As for the real data, fitting the mock data with a model without screening (green) achieves no noticeable $\Delta\log(\mathcal{L})$ for any $\lambda_C$, indicating either that the $\vec{a}_5$ fields of the two models are poorly correlated even at the same $\lambda_C$, or, more likely, that the inference depends crucially on the test galaxies' unscreened likelihood fractions $f$. In the model without screening $f\equiv 1$.

These results suggest that $\lambda_C = 1.8 \: \text{Mpc}, \Delta G/G_N = 0.011$ provides the best solution to our inference problem, and we take it as our fiducial model from now on. This is precisely the maximum-likelihood $\lambda_C$ of D18, with best-fit $\Delta G/G$ differing by a factor of only 2.3.

\begin{figure*}
  \subfigure[$\Delta G/G_N$]
  {
    \includegraphics[width=0.317\textwidth]{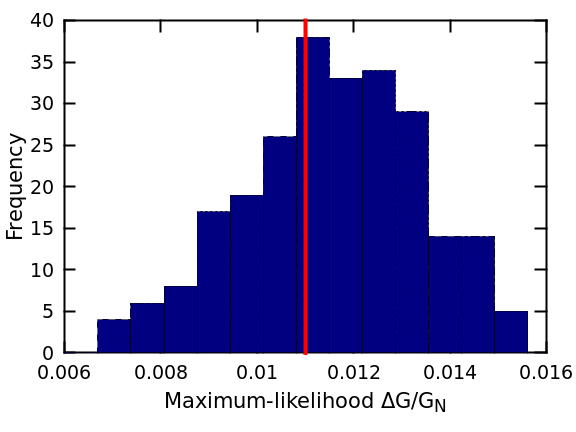}
    \label{fig:mock_dG}
  }
  \subfigure[$\Delta G/G_N-\Delta\log(\mathcal{L})$]
  {
    \includegraphics[width=0.317\textwidth]{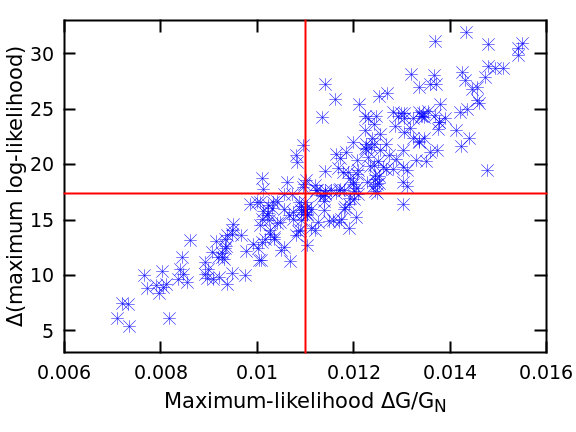}
    \label{fig:mock_dG-dML}
  }  
  \subfigure[$\Delta\log(\mathcal{L})$]
  {
    \includegraphics[width=0.317\textwidth]{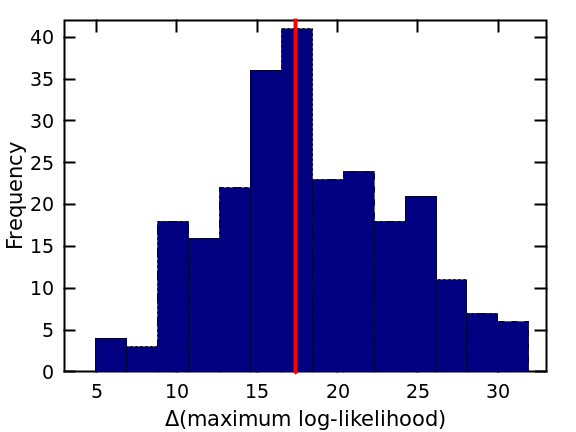}
    \label{fig:mock_dML}
  }  
  \caption{Results of refitting a screened model with $\lambda_C = 1.8$ Mpc to $250$ mock data generated by that model at $\Delta G/G_N = 0.011$, differing only in noise. Fig.~\ref{fig:mock_dG} plots the distribution of best-fit $\Delta G/G_N$ values and Fig.~\ref{fig:mock_dML} the corresponding distribution of $\Delta\log(\mathcal{L})$. The vertical red lines show the values in the real data ($0.011$ and $17.4$ respectively); that these are near the centres of the mock data distributions shows that data to behave very much like the observations. Fig.~\ref{fig:mock_dG-dML} illustrates the correlation between $\Delta G/G_N$ and $\Delta\log(\mathcal{L})$ among the mock datasets, with the real data located at the intersection of the red lines.}
  \label{fig:mocks}
\end{figure*}

In Fig.~\ref{fig:mocks} we generate and refit $250$ mock datasets with the fiducial model to examine the scatter between them. Fig.~\ref{fig:mock_dG} shows the distribution of maximum-likelihood $\Delta G/G_N$: that the reconstructed values cluster around the input value indicates that we are able to reconstruct a known truth without bias. Fig.~\ref{fig:mock_dML} shows the corresponding distribution of $\Delta\log(\mathcal{L})$, demonstrating the signal to be as strong relative to the noise in the mock data as the real data and hence indicating reliability of our noise model. Finally, Fig.~\ref{fig:mock_dG-dML} correlates $\Delta G/G_N$ and $\Delta\log(\mathcal{L})$, showing that mock datasets with a larger best-fit $\Delta G/G_N$ also achieve a larger $\Delta\log(\mathcal{L})$. The real data lies at the intersection of the red lines. Through the lens of our likelihood, mock data with $\lambda_C = 1.8 \: \text{Mpc}$ and $\Delta G/G_N = 0.011$ appears identical to the observations.

Finally, we check the $\Delta\log(\mathcal{L})$ values derived by refitting any model to mock data generated with $\Delta G=0$. We find these to be $\lesssim1$, showing that the peaks in Fig.~\ref{fig:like_like} could not follow from data of this form.

\subsubsection{Bootstrap and jackknife resampling}

\begin{figure*}
  \subfigure[Bootstrap $\Delta G/G_N$]
  {
    \includegraphics[width=0.317\textwidth]{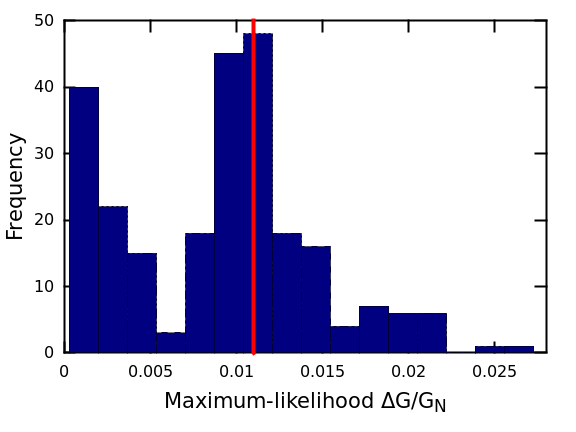}
    \label{fig:boot_dG}
  }
  \subfigure[Bootstrap $\Delta G/G_N-\sigma$]
  {
    \includegraphics[width=0.317\textwidth]{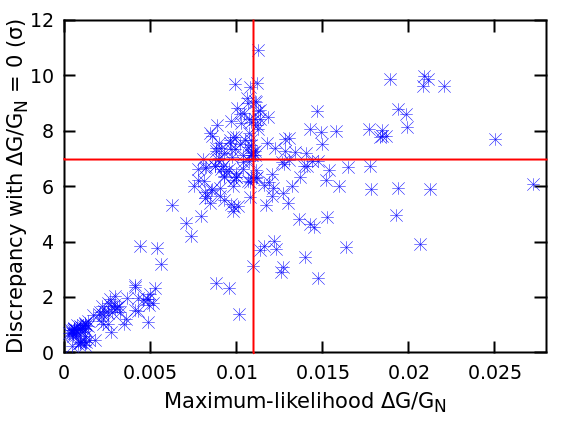}
    \label{fig:boot_dG-dML}
  }
  \subfigure[Bootstrap $\sigma$]
  {
    \includegraphics[width=0.317\textwidth]{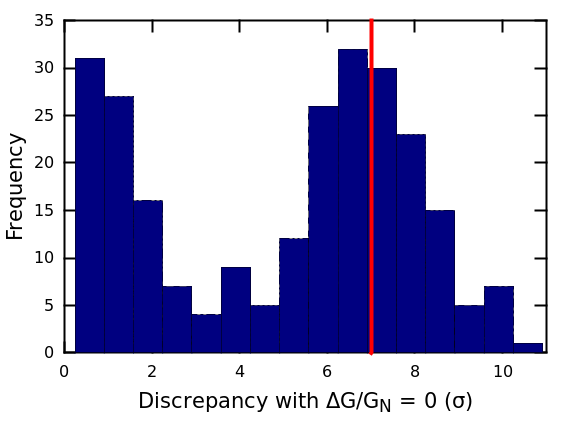}
    \label{fig:boot}
  }
  \subfigure[Jackknife $\Delta G/G_N$]
  {
    \includegraphics[width=0.317\textwidth]{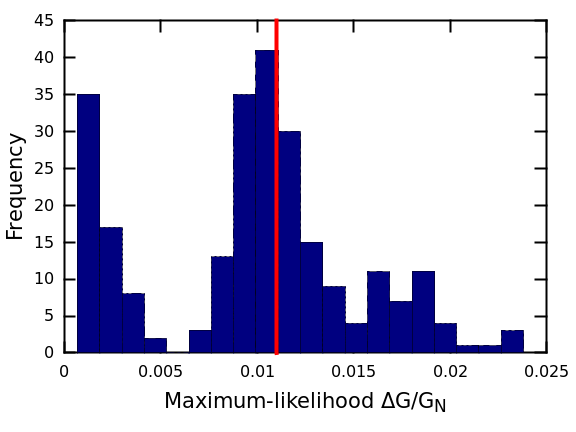}
    \label{fig:jack_dG}
  }
  \subfigure[Jackknife $\Delta G/G_N-\sigma$]
  {
    \includegraphics[width=0.317\textwidth]{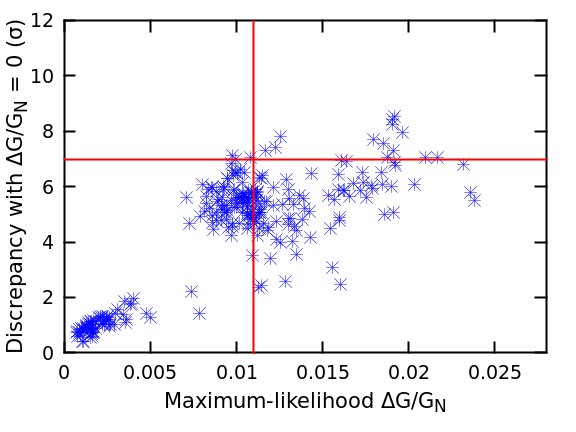}
    \label{fig:jack_dG-dML}
  }
  \subfigure[Jackknife $\sigma$]
  {
    \includegraphics[width=0.317\textwidth]{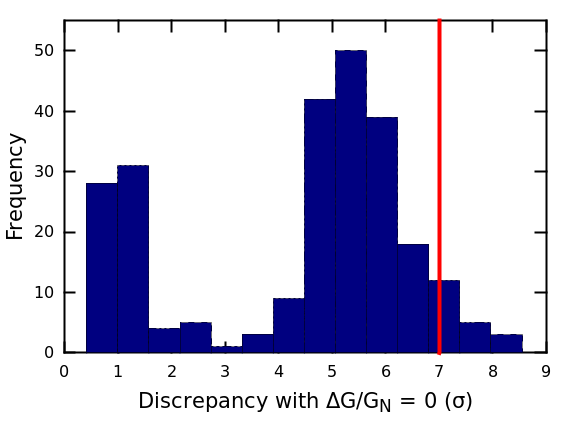}
    \label{fig:jack}
  }
  \caption{Figs.~\ref{fig:boot_dG} \& \ref{fig:jack_dG} show the distributions of maximum-likelihood $\Delta G/G_N$ values (for the screened model with $\lambda_C = 1.8$ Mpc) inferred from 250 bootstrap (upper) or jackknife (lower) resamples of the NSA data. Each jackknife dataset contains $70$\% of the full sample. The vertical red line shows the result in the full sample ($0.011$), around which the resample values cluster. Figs.~\ref{fig:boot} \& \ref{fig:jack} show the corresponding distributions of deviations from $\Delta G=0$, measured by dividing the median of the $\Delta G/G_N$ posterior by its standard deviation. Although both sets produce $\Delta G>0$ with lower significance than the full dataset on the whole, that the majority of resamples achieve a $\gtrsim3\sigma$ result indicates that the preference for a fifth force is not a peculiar property of the NSA sample. Finally, Figs.~\ref{fig:boot_dG-dML} \& \ref{fig:jack_dG-dML} show the correlation between best-fit $\Delta G/G_N$ and discrepancy from $\Delta G=0$. The real data lies at the intersection of the red lines.}  
  \label{fig:boot_jack}
\end{figure*}

Next, we generate $250$ mock datasets by bootstrap- or jackknife-resampling the real data (in the jackknife case retaining a random $70\%$ of the sample). We fit each dataset with the screened $\lambda_C = 1.8$ Mpc model and calculate both the maximum-likelihood $\Delta G/G_N$ and the discrepancy from $\Delta G=0$ exhibited by the posterior, defined as the ratio of the median $\Delta G/G_N$ to the standard deviation. We show the resulting histograms in Fig.~\ref{fig:boot_jack}, where the red lines show the corresponding values in the real data. Most bootstraps have a very similar significance of $\Delta G > 0$ to the real data and most jackknifes a slightly lower significance due to the reduced statistics. There is however a sizeable fraction of datasets in each case with much lower significance, $<$3$\sigma$. Thus in some samples similar to ours a significant deviation from $\Delta G = 0$ would not be found. The maximum-likelihood $\Delta G/G_N$ values tell a similar story: while resamples in which a significant improvement over GR is possible favour $\Delta G/G_N$ values close to that in the total data, the remainder favour $\Delta G \simeq 0$. Of all the checks we have performed this one casts the most doubt on the fifth-force interpretation, and indicates that more data may be required to validate the model with high statistical confidence. That our results here are more ambiguous than the corresponding fig. 12 of D18 likely follows from our smaller sample size.

\subsubsection{Inversion of the predicted signal}

\begin{figure}
  \centering
  \includegraphics[width=0.5\textwidth]{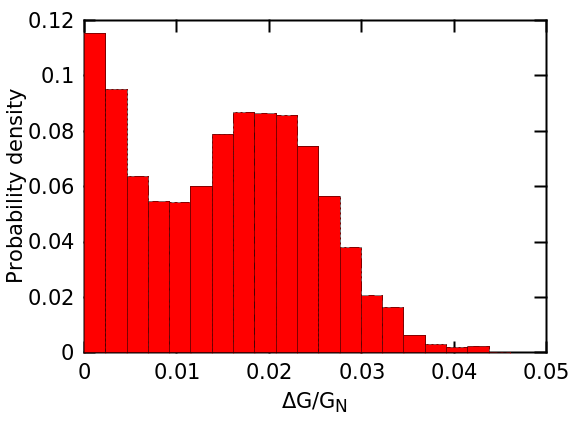}
  \caption{Posterior of $\Delta G/G_N$ for the screened model with $\lambda_C = 1.8$ Mpc, but with the sign of the predicted $w_1$ reversed (c.f. Fig.~\ref{fig:dG3}).}
  \label{fig:reverse}
\end{figure}

As a final check, we repeat the inference at $\lambda_C = 1.8$ Mpc with the sign of all predicted $w_1$ values reversed. The posterior of $\Delta G/G_N$ is shown in Fig.~\ref{fig:reverse} (c.f. Fig.~\ref{fig:dG3}): that the detection goes away shows that the directions of the measured and predicted warps are on the whole aligned on a galaxy-by-galaxy basis when $w_1$ has the correct sign, as required for a component of each warp to derive from $\vec{a}_5$.

\vspace{4mm}

\noindent Although the present analysis uses fewer galaxies than D18 (4,206 vs 10,822), the smaller uncertainty both in the measurements and theoretical prediction (depending on the halo density out to $\sim r_s$, which is better known than in the central $\sim 100$ pc) allows it to achieve a slightly stronger result. This is quantified both by the maximum $\Delta\log(\mathcal{L})$ (20.2 vs 15.6), and the significance of the $\Delta G/G_N$ posterior from 0 at the maximum-likelihood $\lambda_C$ ($7.0\sigma$ vs $6.6\sigma$). Further, if the scatter in $\Delta\log(\mathcal{L})$ across $\lambda_C$ for the unscreened model is indicative of its uncertainty, $\Delta\log(\mathcal{L})$ is known here to $\pm1$ (Fig.~\ref{fig:like_like}), while in D18 it is known only to $\pm4$ (their fig. 8a). Finally, the width of primary peak in $\Delta\log(\mathcal{L})$ is narrower here than in D18, providing a tighter constraint on $\lambda_C$. The present inference seems somewhat cleaner than D18's.

\section{Discussion}
\label{sec:discussion}

\subsection{Systematics}
\label{sec:systematics}

Our principal systematics are as follows. Many are directly analogous to those of D18 sec. VIB, where further details may be found.

We assume that the component of the warps due to anything other than a fifth force can be described by a Gaussian with width dependent only on apparent axis ratio $b/a$. This describes the appearance of warps on the sky rather than their intrinsic properties. Although the astrophysical processes governing warp formation are likely to correlate warp strength with a number of internal and environmental galaxy properties, the exact nature of these correlations is unknown due to small sample sizes in previous warp studies and uncertainty in the underlying theory. Unless these correlations are similar to those induced by fifth forces, however, they would not be expected to significantly bias our inference. This seems unlikely due to the specificity of the dependence of predicted $w_1$ on halo properties $n$, $\rho_{rs}$ and $r_s$ and galaxy size $R_\text{eff}$ in Eq.~\ref{eq:w1_final}, as well as the surrounding matter that sets $\Phi_\text{ex}$ and $\vec{a}_5$. Further, the U-shaped warps expected from fifth forces and to which we tailor our summary statistic are observationally less common than S-shapes, and are therefore less likely to receive large contributions from competing astrophysical effects.

We see two ways in which the impact of non-fifth-force physics may be assessed. The first is to measure or derive a set of internal and external properties for a statistical sample of galaxies that may be expected to correlate with warp strength. By calculating the covariances between $w_1$ (or any other summary statistic) and each of these properties, the dependencies of galaxy warping may be empirically determined. This full covariance matrix may then replace the simple $b/a$-dependent variance in the non-fifth-force part of the likelihood function, allowing one to check that a significant correlation with $\vec{a}_5$ remains even when the noise model incorporates finer-grained correlations. We discuss this further in Sec.~\ref{sec:others}. The second is to measure warp strengths in a population of galaxies produced by hydrodynamical simulations, which in principle incorporate all relevant baryonic physics in $\Lambda$CDM. It is this population that the non-fifth-force part of the likelihood function ought to describe. Unfortunately this requires $\mathcal{O}(10$ pc) spatial resolution in the central regions of halos, which is currently possible only in zoom simulations with poor statistics.

Our determination of the $\Phi$ and $\vec{a}_5$ fields, and the galaxy--halo connection, assume a concordance $\Lambda$CDM cosmology, which is inconsistent in detail with that implied by our fifth-force models (e.g.~\citep{Shi, Fontanot, Lombriser_2, Lombriser_rev}). Nevertheless, the weakness of the fifth force in our preferred scenarios ($\mathcal{O}(1\%)$ that of gravity, with a range $\mathcal{O}(1$ Mpc)) makes changes to the growth rate of structure and abundance and profiles of halos small. Ultimately it will be preferable to repeat our analysis self-consistently with modified gravity simulations~\cite{Puchwein, Mead}.

Following \cite{Jain_Vanderplas}, we neglected the self-gravity of the disk in calculating the warp curve given an offset between the halo and stellar mass centres . The action of neighbouring stellar rings provides a restoring force on the disk, reducing the magnitude of the warp in equilibrium. Incorporating this properly requires the sophisticated machinery of potential theory~\cite{Hunter}, but we can provide an order-of-magnitude estimate of its effect by simply summing the forces due to the disk elements in the warp configuration of Sec.~\ref{sec:statistic} and comparing the resultant to $\vec{a}_5$. We show this calculation in Appendix~\ref{appendix}: the upshot is that the disk self-force is $\mathcal{O}(10^{-2})$ times that from the halo, leading to an $\mathcal{O}(1\%)$ reduction in predicted $w_1$. This is almost certainly too small to impact our inference.

Finally, we have assumed that the warp curve Eq.~\ref{eq:z} would be observed at all times, i.e. that warps due to fifth forces are quasi-stable phenomena. This is likely accurate provided that the fifth-force field driving the warp varies on timescales longer than the dynamical time required for the equilibrium of the disk to become established. However, the excitation modes of the disk required for the maintenance of a warp are not fully understood, and it may be possible for the observed warp curve to vary significantly with time even for fixed $\vec{a}_5$. This is likely to be more important when the warp is forced by one or a few individual massive objects, which will induce stronger tidal effects and perturbations to the galaxy's velocity than the case in which $\vec{a}_5$ receives small contribution from many masses within $\lambda_C$. Fortunately, unscreened objects -- which are the only ones to experience a fifth-force warp at all -- fall mostly in the latter class.

\subsection{Comparison with D18}
\label{sec:me}

Fig.~\ref{fig:like} is very similar to D18 fig. 8. Both analyses find $6-7\sigma$ evidence for a chameleon or symmetron-screened fifth force with range $1.3 < \lambda_C/\text{Mpc} < 2.3$ and strength $0.01 \lesssim \Delta G/G_N \lesssim 0.03$. There is a small offset between the maximum-likelihood $\lambda_C$ values of the two analyses -- 1.4 Mpc here vs 1.8 Mpc in D18 -- although it is within the uncertainties. Our analysis of mock data with an injected $\Delta G/G_N$ (Sec.~\ref{sec:valid}) suggests 1.8 Mpc to in fact be the preferred solution here also. The best-fit $\Delta G/G_N$ values are somewhat further apart -- 0.011 here vs 0.025 in D18 at $\lambda_C = 1.8$ Mpc -- and formally discrepant at just over $3\sigma$.

These differences, however, seem trifling compared to the general agreement that we find. This is all the more remarkable given that the two analyses use largely independent samples of galaxies and signals with qualitatively different dependences on galaxy and halo structural parameters. Although the environmental screening ($\Phi_\text{ex}$) and fifth-force acceleration ($\vec{a}_5$) \emph{fields} are common to both analyses, they are evaluated at different points for different samples, making their \emph{values} largely independent. As they are sourced by masses within $\lambda_C\simeq1.8$ Mpc, regions of the field further apart than this are uncorrelated. The total degree of screening depends also on $\Phi_\text{int}$, which is determined differently in the two analyses. Finally, the relevant projections of $\vec{a}_5$ also differ: in D18 $\vec{a}_5$ was projected onto the plane of the sky to yield independent RA and DEC projections of the vector signal $\vec{r}_*$, while here it is projected onto the disk normals, to determine warping, and then onto the plane of sky to set the observed signal $w_1$. We therefore do not believe there to be significant covariance between these two analyses: their results ought to be largely independent.

We caution however that systematic uncertainties may impact the best-fit values of $\Delta G/G_N$ and $\lambda_C$. The largest in either analysis is likely the central halo density $\rho_0$ that appears in D18 eq. 17, and especially its scatter between galaxies. Fig. 16 of that paper shows that varying this scatter between 0.4 and 1.4 dex causes a variation in best-fit $\Delta G/G_N$ of three orders of magnitude, completely dwarfing the factor of 2.3 difference with the result here. The present analysis is much less prone to uncertainties of this type because $w_1$ is determined by the entire extent of the disk (few kpc), rather than the very central regions that define $r_*$ (few $\times 10^{-2}$ kpc). This is manifest in the appearance of $\rho_{rs}$, the dark matter density at the halo scale radius, in the formula for expected $w_1$ where $\rho_0$ appears in the formula for $\vec{r}_*$. Thus we are able to determine the scatter in the restoring force due to the dark matter self-consistently by modelling the full galaxy--halo connection with AM, rather than having to impose it by hand. The approximate agreement in the reconstructed $\Delta G/G_N$ between the two analyses may therefore indicate that the scatter in $\rho_0$ among galaxies of the same central surface baryonic mass density is indeed $\sim1$ dex, as we assumed in D18. Intriguingly, under a unified model for both warps and gas--star offsets -- which may or may not invoke a screened fifth force -- the consistency of the signals could provide a new handle on halo properties, potentially enabling them to be determined with much greater precision than is currently possible.

While the inferred $\Delta G/G_N$ is sensitive to the average magnitude of the predicted signal at fixed $\Delta G/G_N$, the inferred $\lambda_C$ is sensitive only to the variation of $\Phi$ and $\vec{a}_5$ with the radius from a test point out to which the contributions from masses are summed. In particular, $\lambda_C$ is almost fully insensitive to the detailed distribution of dark matter mass within halos. We therefore believe our constraints on $\lambda_C$ to be significantly more robust than those on $\Delta G/G_N$, and consider similarity between the $\lambda_C-\Delta\log(\mathcal{L})$ relations here and in D18 a more important indicator of the general consistency of the analyses than similarity between the maximum-likelihood $\Delta G/G_N$ values. Although we cannot conclude that a screened fifth force must be responsible for the signals, we can be confident that whatever is responsible operates on a scale of $\sim1.8$ Mpc and determines the signals through a feature common to both, most likely a separation of galaxy mass components.

\subsection{Previous work and the path ahead}
\label{sec:others}

The astronomical study of galaxy warps has been ongoing for several decades (see~\cite{Binney} and~\cite{review2} for reviews). The warp in 21 cm emission of the Milky Way has been known since the 50s~\cite{Burke, Kerr}, and warps of other nearby massive galaxies since the 70s~\cite{Rogstad, Newton}. By the 90s sample sizes were large enough to conclude that the majority of galaxies' H\textsc{i} disks were probably warped~\cite{Bosma}. Evidence for warping of stellar disks (such as we investigate here), however, has proven harder to accumulate as optical emission peters out at a smaller galactocentric radius than H\textsc{i}. Statistical samples of warps in optical disks have been compiled only more recently~\cite{Combes_warps}. The degree of warping has typically been investigated qualitatively by means of visual inspection, although automated algorithms along the lines of ours have previously been employed~\cite{Jim}.

Despite empirical evidence for the prevalence of warps, the processes which drive warp formation are not well understood. In particular, there remains considerable uncertainty on whether warps are generated primarily by factors internal to galaxies or rather by environmental influences~\cite{Combes_warps, correlations}. Those in the latter class include tidal interactions between galaxies, intergalactic magnetic fields and accretion of gas or dark matter, while those in the former rely on interaction between the disk and halo, perhaps due to a triaxial halo misaligned with the disk. Although most warped disks have neighbours massive enough to induce strong tidal fields, warps have also been observed in apparently isolated systems~\cite{Sancisi}. The lifetime of warps is also unclear: while galaxies unlikely have discrete warping modes, they may have long-lived collective excitations resembling warps that are maintained by infall or other interaction with the environment \cite{Sparke}. We can provide only a small piece of information on these puzzles here: that our models without screening achieve no increase in $\log(\mathcal{L})$ over the case $\Delta G=0$ for any $\{\lambda_C, \: \Delta G/G_N\}$ indicates that warps are not significantly correlated with the local acceleration field sourced by matter within any distance from 0.4 to 50 Mpc.

The general mechanisms underlying warp formation have ramifications for our inference: the ``true'' (effective) model of astrophysical warps should determine the non-fifth-force part of our likelihood function. Exactly analogously to D18, the Gaussian component that we convolve with $\mathcal{L}_5$ describes both these additional physical effects and observational uncertainty in the determination of $w_1$. Our assumption of a Gaussian (with width dependent only on $b/a$) will be more accurate the greater the relative importance of the observational uncertainty, which it is better suited to describe. We cannot estimate these uncertainties here, requiring us to extract the information directly from the data.

While most astrophysical processes are expected to generate S- (or integral-) rather than U-shaped warps (and indeed these are more common observationally) \cite{Binney,Garcia}, they almost certainly generate a U-shaped component with strength that correlates significantly with other galaxy properties. In principle it is possible to use our determinations of warp and galaxy properties to investigate this issue and account in part for additional correlations. For example, simply correlating $w_1$ with a number of internal and environmental galaxy properties such as we have already estimated would provide a handle on which are most important. The internal properties are the observed galaxy size, mass and rotation velocity, and the halo mass and concentration from AM; the external properties are the Newtonian potential, acceleration, curvature and ambient density from our gravitational maps. Lack of significant correlation with these properties would indicate either that variations in the warp strength are driven mainly be observational error, or that the astrophysical processes driving warp formation are largely stochastic. In either case our likelihood model for the noise would be reliable. Implementing the remaining correlations in the noise model would give a more accurate baseline on which to add the fifth-force signal.

In the context of modified gravity, warps have been investigated previously in~\cite{Vikram}, using 495 galaxies from SDSS and ALFALFA. These authors were unable to place significant constraints due to the relatively small sample size and inability to forward-model the warp statistic,  but roughly reach sensitivity to $\lambda_C/\text{Mpc} \sim$ few, $\Delta G/G_N \sim$ few. They forecast that $\sim8000$ isolated dwarf galaxies would be required to test $f_{R0}$ to the $2 \times 10^{-7}$ level, i.e. $\Delta G/G_N = 1/3$, $\lambda_C = 1.4$ Mpc. Due to advances in the screening maps, likelihood function and inference machinery, we achieve over an order of magnitude greater sensitivity to $\Delta G/G_N$ at that $\lambda_C$ with only $\sim4000$ galaxies, not all of which are isolated dwarfs. Our inferences greatly surpass in strength all previous fifth-force searches on either astrophysical \cite{Vikram, Vikram_RC, Davis, Sakstein} or cosmological \cite{Lombriser_rev, ISW, ISW_2, Dossett, Smith, RSD, Schmidt, Ferraro, Cataneo, Lombriser, Terukina, Wilcox} scales, and the signal they favour has not been constrained by any previous test.

Besides investigating our systematics further -- including possible contamination from galaxy formation physics -- we plan to extend our framework to the remaining intra-galaxy signals identified in~\cite{Hui, Jain_Vanderplas}: offsets and asymmetries in the rotation curves of stars and gas \cite{Vikram, Vikram_RC}. It is possible to make predictions for these from the best-fit models here and in D18 by means of either the maximum-likelihood estimator or Bayesian posterior predictive distribution: verification of this would provide stronger evidence than the reconstructions presented thus far. Ideally one would infer the model parameters from all of these datasets jointly. We will also improve the resolution of our gravitational maps with constrained N-body simulations using the initial conditions inferred by BORG. Finally, repeating the analyses with independent data of greater quality and quantity in the future -- e.g. from DES, WFIRST, Euclid or LSST -- would provide a further check on the results as well as increase the precision of the inference.

In addition to galaxy formation and dynamics, and fifth forces, warps contain information on possible dark matter interactions which have been proposed to resolve small-scale problems in $\Lambda$CDM \cite{Spergel, SIDM2}. For example, a dark matter self-interaction (SIDM) generates a drag force on a halo as it falls through another halo or the ambient medium, causing a separation between halo and galaxy centres directly analogously to a screened fifth force. This has so far been studied analytically~\cite{Kahlhoefer, Kummer} and in simulations \cite{Secco}, and implies that dark matter self-interactions constitute another systematic for our fifth-force search: the two types of new physics are degenerate in their effect on $w_1$. This is mitigated however by the fact that SIDM would be expected to induce warps pointing predominantly in opposite directions to those from fifth forces, viz. parallel to a galaxy's peculiar velocity rather than opposite the external fifth-force field. That $w_1$ is on the whole anticorrelated with $\vec{a}_5$ as expected from a screened fifth force therefore implies net antialignment with the SIDM prediction and hence the potential for strong constraints on the interaction cross-section $\sigma/m_\text{dm}$. We place these constraints quantitatively in upcoming work (Pardo et al. 2018, in prep).

Regardless of the fate of the detection here and in D18, we foresee galaxy-scale tests of fundamental physics becoming a major frontier of astrophysics in the next decade.

\section{Conclusion}
\label{sec:conc}

With the advent of massive galaxy surveys, intra-galaxy probes are set to become increasingly important in our search for new physics. These are complementary to traditional cosmological probes of $\Lambda$CDM and modified gravity, come at little if any extra experimental cost and reach sensitivities to well-motivated screening mechanisms far surpassing those of any other test. In this paper we search for a chameleon- or symmetron-screened fifth force -- as well as an unscreened fifth force with differential coupling to stars and dark matter -- by investigating the warping of stellar disks that it predicts. This follows from the spatial offset that develops between the stellar disk and halo centre when only one of the two feels the fifth force: in the case of screening this arises when stars self-screen in otherwise unscreened galaxies.

To probe this effect, we reduce $4,206$ $r$-band images of late-type galaxies in the local 100 Mpc from the \textit{Nasa Sloan Atlas} and develop an automated algorithm for quantifying the degree of U-shaped warping on the plane of the sky. We then calculate the expectation for the warp curve of each galaxy as a function of the fifth-force range $\lambda_C$ and strength relative to gravity $\Delta G/G_N$, the screening field $\Phi$ and external fifth-force field $\vec{a}_5$, and the internal structural properties of the galaxies and their halos. By propagating the full non-Gaussian uncertainties in each of these inputs by Monte Carlo sampling we build the fifth-force likelihood function for the observed data $d$, $\mathcal{L}_5(d|\lambda_C, \Delta G/G_N)$, which we convolve with a Gaussian noise component describing both measurement uncertainty and the contribution from non-fifth-force physics. As the expected warp for $\Delta G/G_N$ values of interest, $\mathcal{O}(0.01)$, is only $\mathcal{O}(1-10\%)$ of that measured, the non-fifth-force part of the model accounts for the great majority of the signal. This enables us to set its width equal to the standard deviation of the measured warp in bins of observed axis ratio $b/a$, ensuring that the measurements can be approximately fully accounted for without novel physics. This should make our analysis conservative.

Feeding the data into our likelihood framework and applying Bayes' theorem, we constrain $\lambda_C$ and $\Delta G/G_N$ by Markov Chain Monte Carlo. The overall likelihood is significantly increased ($\Delta\log(\mathcal{L}) \simeq 20$) over the General Relativistic case $\Delta G=0$ by adding a screened fifth force with $\lambda_C \simeq 1.8$ Mpc, $\Delta G/G_N \simeq 0.01$. At the maximum-likelihood $\lambda_C$, the $\Delta G/G_N$ posterior is inconsistent with 0 at $7\sigma$. Neighbouring $\lambda_C$ values exhibit a coherent trend in $\Delta\log(\mathcal{L})$, precisely as would be expected in a fifth-force scenario due to the correlation of $\vec{a}_5$ and $\Phi$ with fifth-force range. Conversely, models of any fifth-force range but without screening achieve no significant increase in likelihood over GR, again as expected from the correlations inherent to the screened model. We check that mock data without a fifth force could not generate our $\{\lambda_C, \Delta G/G_N\}$ posteriors, that the $\Delta\log(\mathcal{L})$ from the real data is compatible at all $\lambda_C$ with that from mock data generated using the best-fit model, that bootstrap- and jackknife-resamples yield similar results to the full dataset, and that the preference for $\Delta G>0$ is eliminated by rotating the predicted warp through $180\degree$.

These results are very similar to those obtained with largely independent data and a completely different signal in \cite{Desmond_PRD}. We cannot be sure that these results could not arise from ``galaxy formation'' physics in $\Lambda$CDM, but we can be confident that some physical process, with a preferred scale and specific correlation with gravitational environment, is separating galaxies' stellar, gas and dark matter mass components. At the very least, our analyses demonstrate the immense and largely undiscovered power of intra-galaxy signals for probing fundamental physics.

\section*{Acknowledgements}

This work was supported by St John's College, Oxford. PGF acknowledges support from Leverhulme, STFC, BIPAC and the ERC. GL acknowledges support by the ANR grant number ANR-16-CE23-0002 and from the Labex ILP (reference ANR-10-LABX-63) part of the Idex SUPER (ANR-11-IDEX-0004-02). We thank James Binney and Kris Pardo for helpful discussions on warps. Computations were performed at Oxford and SLAC.

\newpage

\appendix

\section{The effect of disk self-gravity}
\label{appendix}

In deriving $w_1$ expected from a fifth force in Sec.~\ref{sec:statistic} (Eq.~\ref{eq:w1_final}), we assumed that the only restoring force on the stellar disk came from the halo. In reality, however, each stellar ring interacts with its neighbours to provide an additional force that damps the warp. In this Appendix we estimate the fractional contribution to the overall restoring force from the disk in the warp configuration of Eq.~\ref{eq:z}. If this is small -- as we shall find it to be -- then the perturbation to $w_1$ due to the disk is not important.

Take a point on the disk at $(x_0, y_0, z_0)$ and calculate the force in the $z$ direction due to the rest of the disk. The acceleration from an area element $dA$ at $(x,y,z)$ is
\begin{equation}
da_z = \frac{G_N \: \Sigma(x,y,z) \: dA \: (z-z_0)}{((x-x_0)^2 + (y-y_0)^2 + (z-z_0)^2)^{3/2}}.
\end{equation}
where $\Sigma$ is the surface stellar mass density. As there is azimuthal symmetry around the $z$-axis, we can without loss of generality take $y_0=0$ and specify a general point on the disk with cylindrical polar coordinates ($r,\phi$):
\begin{equation}
x = r \cos(\phi), \; \; \; \; y=r \sin(\phi), \; \; \; \; z = K r^n.
\end{equation}
We also have $dA = r \: d\phi \: dl$, where $dl$ is an infinitesimal displacement along the disk:
\begin{equation}
dl = (dr^2 + dz^2)^{1/2} = dr \: (1 + K^2 n^2 r^{2(n-1)})^{1/2}.
\end{equation}
Using that $\Sigma$ is independent of $\phi$ by symmetry, the total disk self-force in the $z$-direction is then
\begin{align} \label{eq:I}
&a_z = G_N \: K \: \int^{\infty}_0 dr \int^{2\pi}_0 d\phi \: \times \\ \nonumber
&\frac{r \: \Sigma(r) \: (r^n - r_0^n) \: (1 + K^2 n^2 r^{2(n-1)})^{1/2}}{((r \cos(\phi) - x_0)^2 + r^2 \sin(\phi)^2 + K^{2n}(r^n - r_0^n)^2)^{3/2}}.
\end{align}

We now make two simplifications that maintain full generality. First we take the ratio, $f_a$, of $a_z$ with the fifth-force acceleration $a_{5,z}$ that generates the warp. Second, we express $K$ in terms of $w_1$ using Eq.~\ref{eq:w1}:
\begin{equation}
K G_N/a_{5,z} = \frac{3-n}{4 \pi \rho_{rs} r_s^n} = w_1 \frac{(n+1)(n+2)}{n} \: a_{5,z}^{-1} \: (3 R_\text{eff})^{1-n}.
\end{equation}
Thus
\begin{equation} \label{eq:f}
f_a = w_1 \: G_N \: \frac{(n+1)(n+2)}{n} \: a_{5,z}^{-1} \: (3 R_\text{eff})^{1-n} \: I(n, r_0)
\end{equation}
where $I$ is the integral over $r$ and $\phi$ in Eq.~\ref{eq:I}.

From Eq.~\ref{eq:f} the order of magnitude of $f_a$ is already apparent on dimensional grounds. As $I$ has units of $M L^{n-3}$, and the only dimensions in the integral are $M_*$ and $R_\text{eff}$ (which sets $\Sigma$), we must have $I = A M_* R_\text{eff}^{n-3}$ where $A$ is a constant of proportionality. Thus dimensionally
\begin{equation}\label{eq:f_a}
f_a \sim w_1 \: \frac{M_*}{R_\text{eff}^2} \: \frac{G_N}{a_{5,z}}.
\end{equation}
Plugging in some characteristic values for our galaxies -- $w_1 = 10^{-4}$, $M_* = 10^9 M_\odot$, $R_\text{eff}=3 \: \text{kpc}$, $a_{5,z} = 10^{-16}$ km s$^{-2}$ and $G = 4.5 \times 10^{-39} \: \text{kpc}^3 \: \text{s}^{-2} \: M_\odot^{-1}$ -- we find $f \sim 10^{-2}$. This is the approximate factor by which the warp curve $z(r)$ is flattened by the self-gravity of the disk: the change is likely insignificant.

A more precise solution is obtained by evaluating Eq.~\ref{eq:I}. Given that the warp is small, we can expand the integral in powers of $z/r$: the leading-order contribution is when all $K$-dependent terms are dropped. Assuming an exponential disk
\begin{equation}
\Sigma(r) = \frac{M_*}{2 \pi R_\text{d}^2} \exp(-r/R_\text{d}),
\end{equation}
we find
\begin{align}
&I(n,r_0) = \frac{M_*}{2 \pi R_\text{d}^2} \: \int^{\infty}_0 dr \: \int^{2\pi}_0 d\phi \: \times \\ \nonumber
r& \exp(-r/R_\text{d}) (r^n - r_0^n) ((r\cos(\phi) - r_0)^2 + r^2 \sin^2(\phi))^{-3/2}.
\end{align}
The largest value will arise for $r_0 = 0$ (the centre of the disk), since in that case all of the disk is at larger $z$ than the point under consideration. This will therefore give the largest value of $f_a$ and the most stringent test of the approximation that the disk self-gravity is negligible. In this case the $\phi$ integral trivially evaluates to $2\pi$ and we are left with:
\begin{equation}
I(n,0) = \frac{M_*}{R_\text{d}^2} \int^{\infty}_0 dr \: \exp(-r/R_\text{d}) \: r^{n-2}.
\end{equation}
E.g. for $n=1.5$, using that $R_\text{eff} = 1.68 \: R_\text{d}$, we find $I = 3.87 \: R_\text{eff}^{-1.5}$. Thus
\begin{equation}
f_a = 13 \: w_1 \: \frac{M_*}{R_\text{eff}^2} \: \frac{G_N}{a_{5,z}}.
\end{equation}

\end{document}